\documentclass[a4paper,11pt,feynmf]{article} 
\usepackage{amssymb}
\usepackage{amsmath,amssymb,mathrsfs}
\usepackage{graphicx}
\usepackage{subcaption}
\usepackage[numbers,sort&compress]{natbib}
\usepackage{ifpdf}
\usepackage{xcolor}
\usepackage{float}
\usepackage[top=30mm, right=30mm, bottom=25mm, left=25mm]{geometry}
\newcommand*{\rom}[1]{\expandafter\@slowromancap\romannumeral #1@}
\begin{document}

\vskip .3in {\textbf{{Analytical solution of magneto-hydrodynamics with acceleration effects of Bjorken expansion in  heavy-ion collisions }}}
\medskip
\vskip .3in \centerline{  A. F. Kord$^1$, A. Ghaani $^{1}$, M. Haddadi Moghaddam$^{2}$ }
\bigskip
{\small{\it {$^1$ Department of Physics, Hakim Sabzevari University (HSU), P.O.Box 397, Sabzevar, Iran.

$^2$ Department of Physics, University of Turin and INFN, Turin, Via P. Giuria 1, I-10125 Turin, Italy. }}}


\small{\it\center{ E-mail: $afarzaneh@hsu.ac.ir$}} \vskip .3in

\begin{abstract}
In this work, we study the 1+1 longitudinal acceleration expansion of hot and dense quark matter  as  a conducting  relativistic fluid with the  electric conductivity $\sigma$. The plasma has embedded in the presence of electric and magnetic fields which are perpendicular together in the transverse plane. In order to be more realistic, we generalize the Bjorken solution, which includes the acceleration effects on the fluid expansion. We apply a perturbation fashion in initial condition to solve the  relativistic magneto-hydrodynamics  equations. This procedure leads us to achieve the exact algebraic expressions for both electric and magnetic fields. We also find the effects of the electromagnetic fields on the acceleration of the fluid, and the correction of energy density obtained from the magneto-hydrodynamics solutions.  
\end{abstract}


\section{Introduction}

It has been shown that a new form of hot and dense nuclear matter is created in the relativistic heavy-ions collisions, which is commonly known as quark-gluon plasma (QGP). Also, during these collisions, strong electromagnetic fields generated due to the relativistic motion of the colliding heavy ions carrying significant positive electric charges. Recently, there have been some investigations, in a typical non-central collisions such as Pb-Pb at the center of mass energy $ \sqrt{s}=2.76 $ TeV and Au-Au at the center of mass energy $\sqrt{ s }= 200 $ GeV an extremely strong magnetic field is created of the order $ 10^{18}-10^{19} $ G, which corresponds to $ \lvert eB \rvert \simeq 1 $ GeV$^2$ and $10^{13}$ times larger than the effective steady magnetic field which is found in the laboratory \cite{b1,b2,b3,b4,b5,b6}.

It has been argued that the existence of these strong fields is essential for a wide variety of new phenomena such as Chiral Magnetic Effect (CME), Chiral Magnetic wave (CMW), Chiral Electric Separation Effect (CESE), Chiral Hall Separation Effect (CHSE), pressure anisotropy in QGP, influence on the direct and elliptic flow and shift of the critical temperature. Some references and review articles can be found in Refs. \cite{b7,b8,b9,b10,b11,b12,b13,b14,b15,b16}.

Some studies have been done on the space-time evolution of electromagnetic fields created by the colliding charged beams moving at relativistic speed in the z-direction, as a solution of the magneto-hydrodynamic equations \cite{b3,b5,b17,b18,b19,b20,b21}. Additionally, a series of primary results are obtained by estimating the significance of strong EM fields on the QGP medium \cite{b22,b23,b24,b25}. Moreover, some attempts have been made to solve the equations of 1+3 dimensional relativistic magneto-hydrodynamics (RMHD) \cite{b26,b27,b28,b29,b30,b31}. Recently, relativistic magneto-hydrodynamics with longitudinal boost invariance in the presence of finite electric conductivity has also been studied in some works such as \cite{b31a,b31b}.

One of the renowned  pictures which describe the expanding of QGP in the beam direction with the exact solution is the boost-invariant (rapidity-independent) Bjorken picture. This solution predicts a flat rapidity distribution of final particles, which is at variance with observations at RHIC, except for a restricted region around mid-rapidity \cite{b1,b31c}. In realistic collisions, boost invariance is violated, and one should investigate a model that is more faithful and is not far from the accelerationless Bjorken picture. There have been several attempts to generalize the Bjorken model. Some have included accelerating solutions of relativistic fluid dynamics to obtain the more realistic estimation \cite{RD}. 

In this work, we investigate the response of resistive fluid with finite electrical conductivity $\sigma$ to the presence of coupled transverse electric and magnetic fields analytically. Here, we consider the combination of relativistic hydrodynamic equations with Maxwell's equations and solve in 1+1 dimensions a set of coupled RMHD equations. Based on previous studies, we assume a magnetic field is produced in $y$ direction, which is perpendicular to the reaction plane, and the electric field is created in a reaction plane in $x$ direction. We consider the QGP medium has been created a short time after collisions with the  generalized   Bjorken longitudinal expansion. We are interested in obtaining  solutions representing the RRMHD extension of one-dimensional generalized Bjorken flow ($v_z\neq \frac{z}{t}$) along the $z$ direction and  the Lorentz force is directed along $x$ direction.

Outlines of this paper are as follows: In the next section, we introduce the resistive relativistic magneto-hydrodynamic (RRMHD) equations in their most general form, considering them in the case of  plasma with finite electrical conductivity. In Section 3, we solve the RRMHD equations with a perturbative approach and obtain
the analytic solution for the EM fields and the energy density. The discussion on the results and the properties of EM fields, energy density, and acceleration parameter will be presented in Section 4. Finally, the last section is devoted to some concluding remarks.

\section{Resistive relativistic magneto-hydrodynamic}
To describe the interaction of matter and electromagnetic fields in quark-gluon plasma, we consider the relativistic magneto-hydrodynamic (RMHD) framework \cite{b26, b32,b33}. For simplicity, we assume an ideal ultra-relativistic plasma with massless particles and finite electrical conductivity $\sigma$. Thus, in the equation of state, the pressure is merely  proportional to the energy density: $P = \kappa\epsilon$ where $\kappa$
is constant.
For an ideal fluid with finite electrical conductivity which is called resistive fluid,  equations of RRMHD can
be written in the form of  covariant conservation laws:
\begin{eqnarray}
d_\mu T_{matter}^{\mu\nu}&=&-J_\lambda F^{\lambda \nu}\label{1}\\
d_\mu F^{\star\mu\nu}&=&0\label{2}\\
d_\mu F^{\mu\nu}&=&-J^\nu, \qquad\ d_\mu J^\mu=0 \label{3}
\end{eqnarray}
The energy-momentum tensor for the fluid is:
\begin{eqnarray}
T_{matter}^{\mu\nu}=(\epsilon+P)u^\mu u^\nu+Pg^{\mu\nu} , \qquad\ P=\kappa\epsilon
\end{eqnarray}
where $u^\mu$,$ \epsilon$, and $P$ are fluid four-velocity, energy density, and pressure. The metric tensor at flat space-time is $g_{\mu\nu}=diag(-,+,+,+)$, and the  fluid four-velocity ($u_{\mu}u^{\mu}=-1$) is defined as:
$$u^\mu=\gamma(1, \vec v)  , \qquad \gamma=\frac{1}{\sqrt{1-v^2}}$$ 
Also, tensors of the electromagnetic fields and current density are given by:
\begin{eqnarray}
F^{\mu\nu}&=&u^\mu e^\nu-u^\nu e^\mu+\epsilon^{\mu\nu\lambda\kappa}b_\lambda u_\kappa\\
F^{\star\mu\nu}&=&u^\mu b^\nu-u^\nu b^\mu-\epsilon^{\mu\nu\lambda\kappa}e_\lambda u_\kappa\\
J^\mu&=&\sigma e^\mu\label{J}
\end{eqnarray}
Here $\epsilon^{\mu\nu\lambda\kappa}=(-g)^{-1/2}[\mu\nu\lambda\kappa]$ is the Levi-Civita tensor density with $g=\det\{g_{\mu\nu}\}$ and $[\mu\nu\lambda\kappa]$ is the completely antisymmetric Levi-Civita symbol.
Besides:
\begin{eqnarray}
e^\mu=F^{\mu\nu}u_\nu, \qquad \ b^\mu=F^{\star\mu\nu}u_\nu, \ \qquad e^\mu u_\mu=b^\mu u_\mu=0
\end{eqnarray}
$e^\mu$ and $b^{\mu}$ are electric and magnetic field four-vector in the local rest frame of the fluid, which is related to the one measured in the lab frame.

In Eqs. (\ref{1}) to (\ref{3}) the covariant derivatives are given by:
\begin{eqnarray}
d_\mu A^\nu&=&\partial_\mu A^\nu+\Gamma^\nu_{\mu m} A^m\\
d_p A^{\mu\nu}&=&\partial_p A^{\mu\nu}+\Gamma^\mu_{p m} A^{m \nu}+\Gamma^\nu_{p m} A^{ \mu m}
\end{eqnarray}
where $\Gamma^i_{j k}$ are the Christoffel symbols:
\begin{eqnarray}
\Gamma^i_{jk}=\frac{1}{2}g^{im}\left(\frac{\partial g_{mj}}{\partial x^k}+\frac{\partial g_{mk}}
{\partial x^j}-\frac{\partial g_{jk}}{\partial x^m}\right)
\end{eqnarray}
It is more convenient to work with Milne coordinates rather than the standard Cartesian coordinates for a longitudinal boost invariant flow:
\begin{eqnarray}
(\tau, x, y, \eta)=\left(\sqrt{t^2-z^2},x,y,\frac{1}{2}ln\frac{t+z}{t-z}\right)
\end{eqnarray}
Here, the metric is presented as:
\begin{eqnarray}
g^{\mu\nu}=diag(-1, 1, 1, \frac{1}{\tau^2}), \ \ \ \ g_{\mu\nu}=diag(-1, 1, 1, \tau^2)
\end{eqnarray}

By  executing the projection of Eq.~(\ref{1}) along the longitudinal and transverse direction concerning  $u^\mu$, one can rewrite the conservation equations:
\begin{eqnarray}
u_\nu( d_\mu T_{matter}^{\mu\nu}=-J_\lambda F^{\lambda\nu} )
&\rightarrow& D\epsilon+(\epsilon+P)\Theta=e^\lambda J_\lambda \label{e}\\
\Delta_{\alpha \nu} ( d_\mu T_{matter}^{\mu\nu}=-J_\lambda F^{\lambda\nu} )
&\rightarrow&(\epsilon+P)Du_\alpha+\nabla_\alpha P=g_{\alpha \nu} F^{\nu\lambda}J_\lambda-u_\alpha e^\lambda J_\lambda\label{m}\nonumber\\
\end{eqnarray}
where:
\begin{eqnarray}
D=u^\mu d_\mu, \quad \Theta=d_\mu u^\mu,\quad \Delta_{\alpha \nu}=g_{\alpha \nu}+u_{\alpha} u_\nu, \quad \nabla^\mu=d^\mu+u^\mu D
\end{eqnarray}

In contrast with the energy-momentum tensor $T^{\mu\nu}$, the dual electromagnetic tensor $F^{*\mu\nu}$ is
anti-symmetric; hence the homogeneous Maxwell's equation, $d_\mu F^{*\mu\nu}=0$,
leads to the following equations:
\begin{eqnarray}
\partial_x F^{*x\tau}+\partial_y F^{*y\tau}+\partial_\eta F^{*\eta\tau}&=&0\label{M1}\\
\partial_\tau F^{*\tau x}+\partial_y F^{*y x}+\partial_\eta F^{*\eta x}+\frac{1}{\tau}F^{*\tau x}&=&0\label{M2}\\
\partial_\tau F^{*\tau y}+\partial_x F^{*x y}+\partial_\eta F^{*\eta y}+\frac{1}{\tau}F^{*\tau y}&=&0\label{M3}\\
\partial_\tau F^{*\tau \eta}+\partial_x F^{*x\eta}+\partial_y F^{*y \eta}+\frac{1}{\tau}F^{*\tau\eta}&=&0\label{M4}
\end{eqnarray}
Also, the in-homogeneous Maxwell's equations $(d_\mu F^{\mu\nu}=-J^\nu)$ are given by:
\begin{eqnarray}
\partial_x F^{x\tau}+\partial_y F^{y\tau}+\partial_\eta F^{\eta\tau}&=&-J^\tau \label{IM1}\\
\partial_\tau F^{\tau x}+\partial_y F^{y x}+\partial_\eta F^{\eta x}+\frac{1}{\tau}F^{\tau x}&=&-J^x \label{IM2}\\
\partial_\tau F^{\tau y}+\partial_x F^{x y}+\partial_\eta F^{\eta y}+\frac{1}{\tau}F^{\tau y}&=&-J^y \label{IM3}\\
\partial_\tau F^{\tau \eta}+\partial_x F^{x\eta}+\partial_y F^{y \eta}+\frac{1}{\tau}F^{\tau\eta}&=&-J^\eta\label{IM4}
\end{eqnarray}

\section{ Equations with longitudinal acceleration in 1+1 dimensional }
In this section,  we  investigate one-dimensional generalized Bjorken flow ($v_z\neq \frac{z}{t}$) along the $z $ direction with velocity $u^\mu=\gamma(1, 0, 0, v_z)=(\cosh{Y},0,0,\sinh{Y})$. The non-central collisions can create a strong magnetic field outside the reaction plane, which is dominated by a $y$ component. The magnetic field induces the  electric field in the reaction plane, which is dominated by a $x$ component because of  Faraday and Hall effects \cite{b34,b35}. In fact, because  of  electromagnetic fields, there is a   pressure gradient that causes  the boost invariant Bjorken flow condition ($v_z= \frac{z}{t}$)  is broken explicitly. We look for the effects of the EM fields on the  longitudinal  acceleration  and the  expansion of the flow.  

We assume that  the electric field is oriented in $x$ direction and the magnetic field is perpendicular to the reaction plane, pointing along the $y$ direction in an inviscid fluid with finite electrical conductivity:
\begin{eqnarray}\label{em1}
 b^\mu=(0, 0,  b^y, 0),\qquad e^\mu=(0, e^x, 0, 0)
\end{eqnarray}
 
Also, we consider that the fluid  has  only longitudinal flow, and  the transverse flow is neglected. We  parameterize the fluid  four-velocity as follow:
 \begin{eqnarray}
 u^\mu=\gamma(1, 0, 0, v_z)=(\cosh{Y},0,0,\sinh{Y}).  
\end{eqnarray}
that $Y$ is fluid  rapidity( $v_z = \tanh Y$).
The four-velocity in Milne coordinates becomes:
\begin{eqnarray}
u^\mu=(\cosh{(Y-\eta)},0,0,\frac{\sinh(Y-\eta)}{\tau})={\bar{\gamma}}(1, 0, 0,  \frac{\tanh{(Y-\eta)}}{\tau} )={\bar{\gamma}}(1, 0, 0, \frac{\bar{v}}{\tau}) \label{Ini}\nonumber\\
\end{eqnarray}
where $\bar{\gamma}=\cosh(Y-\eta)$, and $\bar{v}=\tanh{(Y-\eta)}$. Thus, one obtains:
\begin{equation}
D=\bar{\gamma}(\partial _{\tau} + \frac{1}{\tau} \bar{v} \partial_\eta) , \quad \Theta= \bar{\gamma} (\bar{v} \partial _{\tau} Y +  \frac{1}{\tau} \partial_\eta Y)
\end{equation}

Regarding  the above setup, for conservation  Eqs. (\ref{e})--(\ref{m}) one can write:	
\begin{eqnarray}
(\tau\partial_\tau+ \bar{v} \partial _{\eta})\epsilon+(\epsilon+P)(\tau \bar{v} \partial _{\tau}Y+\partial_\eta Y)  &=&\frac{1}{\bar{\gamma}}\tau\sigma e_x^2\label{E1}\\
(\epsilon+P)\partial_\tau u^i+\partial^i P&=&0\label{E2}\\
 (\epsilon+P)(\tau\partial_\tau+ \bar{v} \partial _{\eta})Y+(\tau \bar{v} \partial _{\tau}+\partial_\eta )P
 &=& \frac{\sigma\tau}{\bar{\gamma}} e_x b_y\label{E3}
\end{eqnarray}

Further, inserting the initial condition (\ref{em1}) and (\ref{Ini}) into Maxwell's equations (\ref{M1})--(\ref{IM4}) give:
\begin{eqnarray}
\partial_x e_x=\partial_y e_x=\partial_x b_y=\partial_y b_y&=&0\label{E4}\\
\partial_\tau\big[\tau(\bar{\gamma}b_y+\bar{\gamma}\bar{v}e_x )\big] +\partial_\eta \big[\bar{\gamma}\bar{v}b_y+\bar{\gamma}e_x \big]&=&0 \label{E5}\\
\partial_\tau \big[\tau(\bar{\gamma} e_x+\bar{\gamma}\bar{v}b_y)\big] +\partial_\eta \big[(\bar{\gamma}\bar{v} e_x+\bar{\gamma}b_y)\big]&=&- \sigma \tau e_x\label{E6}
\end{eqnarray}
	
\subsection{Perturbative solutions of RRMHD equations  with acceleration}

We now study  perturbative solutions of the conservation equations in the presence  of weak electric and magnetic fields  on the longitudinal expansion of the flow. We suppose the magnitude of  electromagnetic fields  are suppressed by the energy density of the fluid, $\frac{(b_y^2+e_x^2)}{\epsilon}\ll 1$ which is not far from reality~\cite{b1}. The  electromagnetic fields are oriented in the transverse plane and are perpendicular to the longitudinal direction. We assume the  electromagnetic fields only  depend on  the proper time $\tau$ and  rapidity $\eta$    for a fluid  following  the longitudinal expansion  with an acceleration. Therefore, we  perform the following setup in  Milne  coordinates:
\begin{eqnarray}
\qquad b^\mu=(0, 0, \lambda_1 b_y, 0), \qquad e^\mu=(0,\lambda_1 e_x, 0, 0) \nonumber    \\   
\epsilon(\tau, \eta)=\epsilon_0(\tau)+\lambda_1^2  \epsilon_1(\tau, \eta), \qquad \epsilon_0(\tau)=\epsilon_c (\frac{\tau_0}{\tau})^{1+\kappa}\label{we}
\end{eqnarray}
Here $\tau_0$ is an initial time, and  $\epsilon_c$ represents the initial energy density of the medium at $\tau_0$. Also, $\lambda_1$ is an  expansion parameter in calculations which will be set to unity. 

Moreover, we assume:
\begin{eqnarray}
Y(\tau,\eta)=\lambda(\tau,\eta) \eta=(1+\lambda^*(\tau,\eta)) \eta \longrightarrow Y-\eta=\lambda^*(\tau,\eta)\eta
\end{eqnarray}
where $\lambda^*$ is a very small constant ($0 \le \lambda^* \ll 1$) and $\lambda(\tau,\eta)=1+\lambda^*(\tau,\eta)$ is defined as acceleration parameter \cite{RD}. 

Besides,  we consider  $(Y-\eta) \ll 1$, so we have:
 \begin{eqnarray}
&&\bar{\gamma}=\cosh(Y-\eta)\simeq 1 \\
&&\bar{v}=\tanh{(Y-\eta)} \simeq (Y-\eta)=\lambda^*\eta=\lambda_1 ^2 v   \label{la}
\end{eqnarray}
Thus, the four-velocity in Milne  coordinates becomes:
 \begin{eqnarray}
 u^\mu=(\cosh{(Y-\eta)},0,0,\frac{\sinh(Y-\eta)}{\tau})\simeq(1, 0, 0, \lambda_1^2 \frac{v}{\tau}) \label{Ini1}
 \end{eqnarray}

Substituting  this setup in conservation  Eqs.~(\ref{E1})-(\ref{E3}) up to $O(\lambda_1^2)$, we have:
\begin{eqnarray}
\partial_\tau\epsilon_1+\frac{(1+\kappa) \epsilon_1}{\tau} + \frac{(1+\kappa) \tau_0^{\kappa+1} \epsilon_c}{\tau^{\kappa+2}} \partial_\eta v  &=&\sigma e_x^2\label{et}\\
(\epsilon+P)\partial_\tau u^i+\partial^i P&=&0\label{press}\\
\partial_\eta \epsilon_1+\frac{(1+\kappa)\tau_0^{\kappa+1}\epsilon_c}{\kappa\tau^{\kappa}}\partial_\tau v - \frac{(\kappa^2 -1)\tau_0^{\kappa+1}\epsilon_c}{ \kappa\tau^{\kappa+1}}v&=& \frac{\sigma\tau}{\kappa} e_x b_y\label{eeta}
\end{eqnarray}
By considering a uniform pressure in the transverse plane, the second term in Eq. (\ref{press}) will disappear; thus, one realizes that if the velocities in the $x$ and $y$ directions are initially zero, they will remain zero  at a later time. However, because  of  electromagnetic fields, there is a  pressure gradient given in  $\eta$ direction, so Eq.~ (\ref{eeta}) breaks  the Bjorken flow ($v_z= \frac{z}{t}$) condition explicitly as we supposed. In fact,
there is an  acceleration of the fluid by the EM fields.

Also, applying the above initial condition, Maxwell's equations~(\ref{E5}) and (\ref{E6}) up to $O(\lambda_1^2)$ reduce to following forms:
\begin{eqnarray}
\partial_\tau b_y+\frac{1}{\tau}\partial_\eta e_x+\frac{b_y}{\tau}&=&0 \label{w1}\\
\partial_\tau e_x+\frac{1}{\tau}\partial_\eta b_y+\frac{e_x}{\tau}+\sigma e_x&=&0 \label{w2}
\end{eqnarray}
and the combination of two Eqs.~(\ref{et}) and (\ref{eeta}) gives an in-homogeneous equation for $v(\tau,\eta)$:
\begin{eqnarray}\label{uinh}
 (\kappa +1) \tau_0^{\kappa+1} \epsilon _c \tau ^{-\kappa -2} \Big(  \kappa  \partial^2_\eta v  + \tau  (\kappa-2) \partial_\tau v - \tau^2  \partial^2_\tau v \Big) \nonumber\\
 +\sigma  e_x \Big(\tau \partial_\tau b_y -2 \kappa \partial_\eta e_x \Big)+\sigma  b_y \Big((\kappa +2) e_x+\tau  \partial_\tau e_x\Big)=0
\end{eqnarray}
that directs us to the calculation of  $\lambda^*$ and the acceleration parameter.
For this purpose, we first calculate the electromagnetic fields. Using Eqs.~(\ref{w1}) and~(\ref{w2}), one can obtain the following results:
\begin{eqnarray}
-\partial_\eta^2 e_x+\tau^2\partial_\tau^2 e_x+(3\tau+\sigma\tau^2)\partial_\tau e_x+(1+2\sigma\tau)e_x&=&0 \label{ee}\\
-\partial_\eta^2 b_y+\tau^2\partial_\tau^2 b_y+(3\tau+\sigma\tau^2)\partial_\tau b_y+(1+\sigma\tau)b_y&=&0 \label{bb}
\end{eqnarray}

In order to solve the above equations, we  consider the electric and magnetic fields as:
\begin{eqnarray}\label{gs}
e_x(\tau,\eta)=g(\eta) f(\tau)\nonumber\\
b_y(\tau,\eta)=s(\eta) h(\tau)
\end{eqnarray}
First of all, we  discuss the behavior of functions $g(\eta)$ and $s(\eta)$. They can be written as:
\begin{eqnarray}
&&\frac{d^2 g(\eta)}{d\eta^2}=m^2 g(\eta)\label{Si}\\
&&\frac{d^2 s(\eta)}{d\eta^2}=n^2 s(\eta)\label{Co}
\end{eqnarray}
Also, If we substitute Eq. (\ref{gs})  into Eqs. (\ref{w1}) and (\ref{w2}), we have:
\begin{eqnarray}
\tau h'(\tau)+h(\tau)&=&C f(\tau)\\
g'(\eta)&=&C s(\eta)
\end{eqnarray}
and:
\begin{eqnarray}
\tau f'(\tau)+(1+\sigma\tau)f(\tau)&=&C' h(\tau)\\
s'(\eta)&=&C' g(\eta)
\end{eqnarray}
By using the above relations, it is easy to show that $CC'=m^2=n^2$ are separation constants, and  for further application, we choose $m^2=l(l +2)+1$. Thus, the solutions of Eqs. (\ref{Si}) and (\ref{Co}) are given by:
\begin{eqnarray}
g(\eta)&=& \sum _ m e_m \cosh(m\eta)+e^{\prime}_m \sinh(m\eta)\\
s(\eta)&=&  \sum _ m b_m \sinh(m\eta)+b^{\prime}_m \cosh(m\eta) 
\end{eqnarray}
It has realized  that   the magnetic field generated by two nuclei passing each other in heavy-ion collisions should be an even function of $\eta$ and most dominant at considerable rapidity after collisions \cite{b31}. The assumption is based on the external magnetic field led by moving nuclei after their collisions.  Besides, in central rapidity, after the QGP is formed, there is no external electric field left on average. Thus,  it seems the external electric field is an odd  function of $\eta$. However, when considering the dynamical electromagnetic fields in QGP, one can not  apply the same rapidity profile. Roughly  speaking, we may assume that the dynamical electromagnetic fields follow similar patterns as the external ones. Based on the above discussion, in our setup, we expect  the magnetic field is an even function of $\eta$, and the electric field is an odd function of $\eta$; thus:
\begin{eqnarray}
s(\eta)= \sum _ m b^{\prime}_m \cosh(m\eta), \qquad  \qquad  g(\eta)= \sum _ m e_m\sinh(m\eta)
\end{eqnarray}

Let us to solve Eq. (\ref{ee}). The equation is given by:
\begin{eqnarray}
&& \tau^2\frac{d^2 f(\tau)}{d\tau^2}+(3\tau+\sigma\tau^2)\frac{d f(\tau)}{d\tau}+ \big(-l(l+2)+2\sigma\tau \big) f(\tau)=0
\end{eqnarray}

Then, we consider $f(\tau)=\tau^l G(\tau)=\tau^l\sum_{n=0}^{\infty} a_n\tau^n$ is a power series function. When it is substituted into the above differential equation, we find $G(\tau)$ obeys by:
\begin{eqnarray} \label{G}
&& \frac{d^2 G(\tau)}{d\tau^2}+(\sigma+\frac{2l+3}{\tau})\frac{d G(\tau)}{d\tau}+\sigma\frac{(l+2)}{\tau}G(\tau)=0 
\end{eqnarray}
and it's analytical solution is:
\begin{eqnarray}
G(\tau)&=& C_{l1} e^{-\frac{1}{2}\sigma \tau} \tau^{-l-\frac{3}{2}}\left( \sigma \tau I_{-\frac{1}{2}+l}(-\frac{1}{2}\sigma \tau) + (\sigma \tau +4l+2) I_{l+\frac{1}{2}}(-\frac{1}{2}\sigma \tau)\right) + \nonumber\\
&&C_{l2} e^{-\frac{1}{2}\sigma \tau} \tau^{-l-\frac{3}{2}}\left( - \sigma \tau K_{-\frac{1}{2}+l}(-\frac{1}{2}\sigma \tau)+ (\sigma \tau +4l+2) K_{l+\frac{1}{2}}(-\frac{1}{2}\sigma \tau) \right) \nonumber\\ 
\end{eqnarray}
where $I_{\alpha}$ and $K_{\alpha}$ are the modified Bessel functions. The constant coefficients $C_{l1}$ and $C_{l2}$ are  determined by physical conditions. 

We know $f(\tau)$ should be a well-behaved function at infinity; therefore, we can show  that  $l$  must be: $l=-2,-3,...$. Besides, we have chosen $m^2=l(l +2)+1$, so for $l=-2,-3,.. m^2=1,2,3,..$. This means that $m$ is an integer (we consider $m$ be positive).

Equivalently, the solution of $f(\tau)$ for different values of $l$ can be written as follows:
\begin{eqnarray}
&&f_1(\tau)=\frac{G_1(\tau)}{\tau^2}= C_{11} \frac{1}{\tau^2}-C_{12}\frac{ e^{-\sigma\tau}(\sigma\tau+1)}{\tau^2} \nonumber \\ &&
f_2(\tau)=\frac{G_2(\tau)}{\tau^3}= C_{21}\frac{(3-\sigma\tau)}{\tau^3}+C_{22}\frac{ e^{-\sigma\tau}(\sigma^2\tau^2+4\sigma\tau+6)}{\tau^3}  \nonumber \\ &&
f_3(\tau)=\frac{G_3(\tau)}{\tau^4}=C_{31} \frac{(20-8\sigma\tau -8\sigma^2\tau^2)}{\tau^4}-C_{32}\frac{ e^{-\sigma\tau}(\sigma^3\tau^3+9\sigma^2\tau^2+36\sigma\tau+60)}{\tau^4}  \nonumber \\ &&
, ... .  
\end{eqnarray}
Also, we know if $\sigma\rightarrow \infty $, then $e_x$ should tend  to zero. Therefore, we have to set $C_{i1}=0, i=1,2,3,...$. 

Finally, the most general solution for the electric field is:
\begin{eqnarray}
e_x(\tau,\eta)&=&\sum_{m=1}^{\infty}e_m f_m(\tau) \sinh(m\eta) \nonumber\\
&=&e_1\sinh(\eta)\frac{ e^{-\sigma\tau}(1+\sigma\tau)}{\tau^2} +
e_2 \sinh(2\eta)\frac{e^{-\sigma\tau}(6+4\sigma\tau+\sigma^2\tau^2)}{\tau^3} +\nonumber\\
&& e_3\sinh(3\eta)\frac{e^{-\sigma\tau}(60+36\sigma\tau+9\sigma^2\tau^2+\sigma^3\tau^3)}{\tau^4}+... .
\end{eqnarray}
where $e_{m}$ are constants and should be determined. To find the time dependence of the magnetic field, one can   replace the $e_x(\tau,\eta)$ from the above equation into  Eq. (\ref{w2}). Then, the magnetic field is given by:

\begin{eqnarray}\label{b}
b_y(\tau,\eta)&=& F_1(\tau)+ e_1\cosh(\eta) \frac{e^{-\sigma\tau}}{\tau^2} +
e_2 \cosh(2\eta)\frac{e^{-\sigma\tau}(6+2\sigma\tau)}{\tau^3} +\nonumber\\
&& e_3\cosh(3\eta)\frac{e^{-\sigma\tau}(60+24\sigma\tau+ 3\sigma^2\tau^2)}{\tau^4}+... .
\end{eqnarray} 
We also know if $\sigma \rightarrow \infty $ then $b_y$ tends toward $\frac{1}{\tau}$ \cite{b29}. Therefore, we have $F_1(\tau)=b_0 \frac{\tau_0}{\tau}$. On the other hand, it is expected in vacuum  ($\sigma \rightarrow 0 $) the magnetic field decays as \cite{b20,b31a}:
\begin{eqnarray}
b_y(\tau,0) \approx b_1\frac{\tau_0^3}{\tau^3}
\end{eqnarray} 

 By using this relation, one can deduce  that all coefficients of $e_i$ should be zero except $e_2$. Consequently, the electric  and magnetic fields can be written as:
\begin{eqnarray}\label{eb}
&& e_x(\tau,\eta) = b_1\frac{\tau_0^3}{\tau^3} e^{-\sigma\tau} \sinh(2\eta) (6+4\sigma\tau+\sigma^2\tau^2)\nonumber\\
&& b_y(\tau,\eta) = b_0 \frac{\tau_0}{\tau}(1-e^{-\sigma\tau}) + b_1\frac{\tau_0^3}{\tau^3} e^{-\sigma\tau} \cosh(2\eta) (6+2\sigma\tau) \nonumber\\
\end{eqnarray}

In order to evaluate our results, we consider the ideal RMHD, i.e. electrical conductivity is infinite ($\sigma\to\infty$). In this case, we have:
\begin{eqnarray}
 \ \ \ b_y(\tau)= b_0(\frac{\tau_0}{\tau}), \ \ \ e_x=0
\end{eqnarray}
 which is consistent with previous studies. 

To determine the acceleration parameter, we should look into Eq.~(\ref{uinh}), which consists of homogeneous and inhomogeneous solutions. When $ e_x=0$, we reach a homogeneous partial differential equation that is the result of generalized Bjorken flow. According to the Ref.~\cite{b36}, and simplifying our calculation, we assume $\lambda^*$ only depends on the proper time $\tau$ in the absence of electromagnetic fields in hydrodynamic equations:
\begin{eqnarray}
v^h(\tau,\eta)=\lambda^*(\tau) \eta
\end{eqnarray}
Thus, $\partial^2_\eta v^h(\tau,\eta)=0$ in Eq.~(\ref{uinh}), and the regular solution takes the form:
\begin{eqnarray}\label{vh}
v^h(\tau,\eta)= A_0 \: (\frac{\tau_0}{\tau})^{2/3}\eta
\end{eqnarray}
where $\kappa=\frac{1}{3}$ is chosen.

In order to obtain an inhomogeneous solution of Eq.~ (\ref{uinh}), we substitute  EM fields Eq. (\ref{eb}) and solve  this  differential equation. So, $v^{ih} (\tau,\eta)$ is:
\begin{equation}
v^{ih}(\tau,\eta) = \sigma \tau_0 \big(\frac{\tau_0}{\tau}\big)^{11/3}  \,e^{-2\sigma \tau} \,\sinh(2\eta) \,\cosh(2\eta) \,H(x)\label{u},
\end{equation}
here  $x=\sigma \tau$, and $H(x)$ satisfies the following equation:
\begin{eqnarray}
&&2 \epsilon_c \left(-x^2 \text{H}''(x)+(4 x^2 +\frac{17}{3} x ) \text{H}'(x)-(4 x^2 +\frac{34}{3} x +\frac{17 }{3}) \text{H}(x)\right)=\nonumber\\
&& b_1^2  \left(8 x ^4+60 x ^3+199 x ^2+348 x +270\right)
\end{eqnarray}
Because of the electromagnetic fields, we force to consider  $\lambda$ be dependent on proper time $\tau$ and rapidity $\eta$. Then, the acceleration parameter is given by:

\begin{eqnarray}\label{u}
\lambda(\tau,\eta)=1+ A_0 \: (\frac{\tau_0}{\tau})^{2/3}+
 \frac{ \sigma \tau_0 \big(\frac{\tau_0}{\tau}\big)^{11/3}  \,e^{-2\sigma \tau} \,\sinh(2\eta) \,\cosh(2\eta) \,H(x)}{\eta}
\end{eqnarray} 

In the above equation, there are three terms. If the Bjorken model is considered with $e_x=0$, it leads to $\lambda=1$. However, if the Bjorken Model is generalized ($v_z\neq\frac{z}{t}$) with $e_x=0$, then $\lambda=1+A_0 \: (\frac{\tau_0}{\tau})^{2/3} $. Finally, the third term only exists if the plasma has  finite electrical conductivity($e_x\neq0$).

Moreover, we can achieve the modification of energy density $\epsilon_1$  from Eq. (\ref{eeta}). We will explain  the behavior of these physical quantities in the next section.

\section{Results and discussions}

In this section, we provide a clear insight from the dynamical evolution and features of EM fields, which are given by Eq.~(\ref{eb}). Besides, we will present  the longitudinal evolution of the fluid, which manifests itself in the form of an acceleration parameter that is the effect of generalized Bjorken model and considering a finite electrical conductivity in the fluid. The acceleration parameter and the correction of energy density are obtained from our perturbation  approach. These  two quantities will help us in understanding the space-time evolution of the quark-gluon plasma in heavy-ion collisions. 

In order to evaluate the EM  fields from  Eq.~(\ref{eb}), and to fix the constants $b_0$ , $b_1$, we
consider the initial condition for magnetic field at the mid-rapidity as $b(\tau_0,0)=0.0018 \frac{GeV^2}{e}$ \cite{b34}.
Besides, we need to assume a simple relation between these two coefficients for the simplicity of our calculation. In fact,  we define $|\frac{b_1}{b_0}|=\alpha$ throughout this paper, and we discuss on effects of the different value of $\alpha$ on the evolution of QGP. Due to the electric field being directed along the negative (positive) direction at positive (negative) rapidity \cite{b35}, we choose $b_1$ with a negative sign. This choice leads us to select $b_0$ with a negative sign too; thus, the magnetic field pointing along the $-y$ direction. 

At first, we investigated  the dynamical evolution of EM fields. In Fig.~1, for different values of $\alpha$, we plot  the evolution of $b_y(\tau,\eta)$ as a function of $\tau$ at $\eta=0$. These plots show that the presence of  a conducting medium delays  the decay of the magnetic field; however, the $\alpha$ parameter, which comes from initial conditions, has a significant effect on the delay process of $b_y(\tau,\eta)$. Fig.~1(a) reports the evolution of the magnetic field for $\alpha=0.01$, and we can see that our result for  $\sigma=0.023 \ fm^{-1}$ support  previous  results in \cite{b24}. For larger values of $\alpha$ ( Fig.~1(c) and  Fig.~1(d)) the decay behavior of the magnetic field is not sensitive to the variation of $\sigma$; thus we select the order of $\alpha$ as $0.01\sim0.1$ which the effect of electrical conductivity is remarkable, and we can focus on the dependence of physical quantities to electrical conductivity.

\begin{figure}
\begin{subfigure}[b]{0.53\textwidth}
\includegraphics[width=\textwidth]{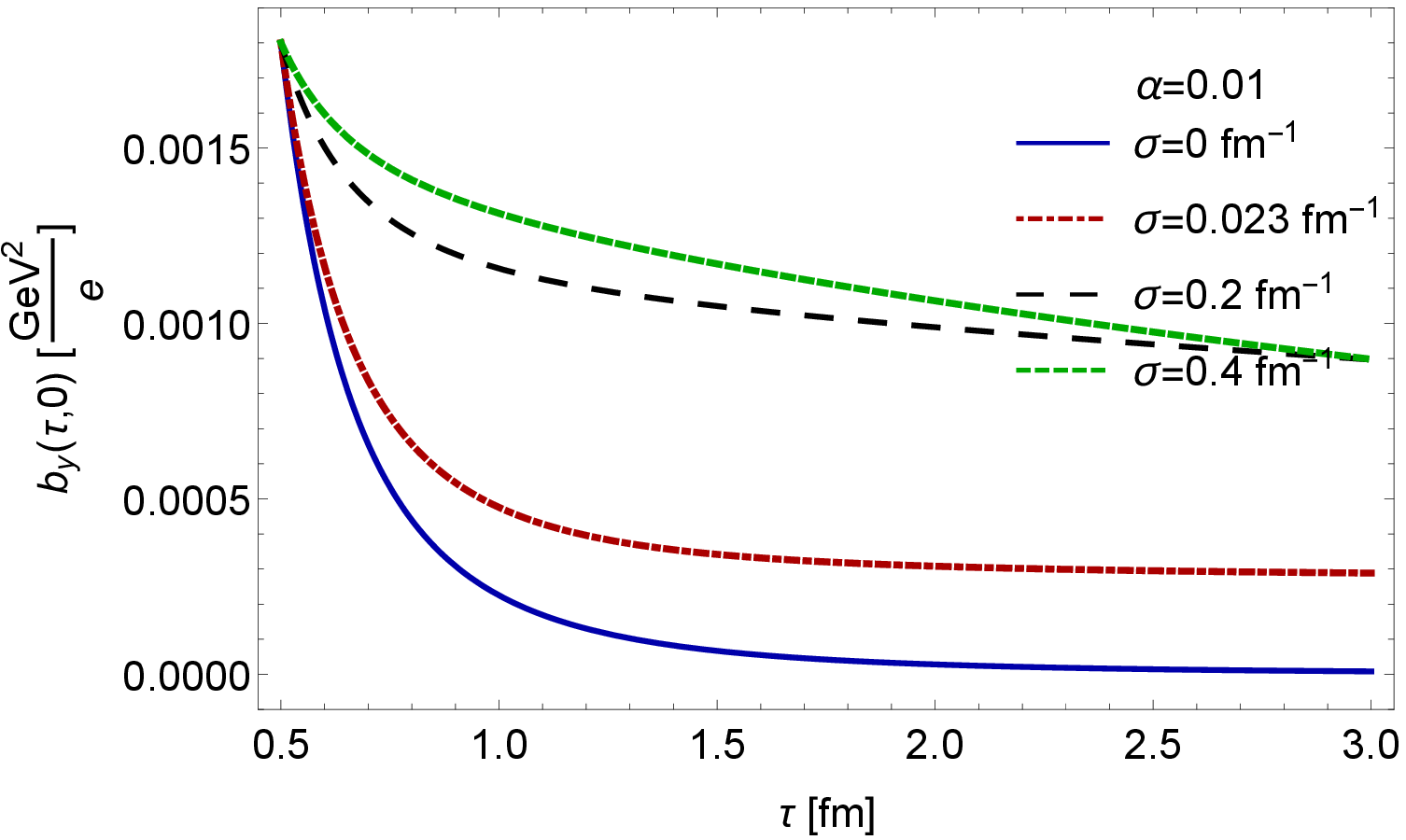}
\caption{}
\end{subfigure}
\hfill
\begin{subfigure}[b]{0.53\textwidth}
\includegraphics[width=\textwidth]{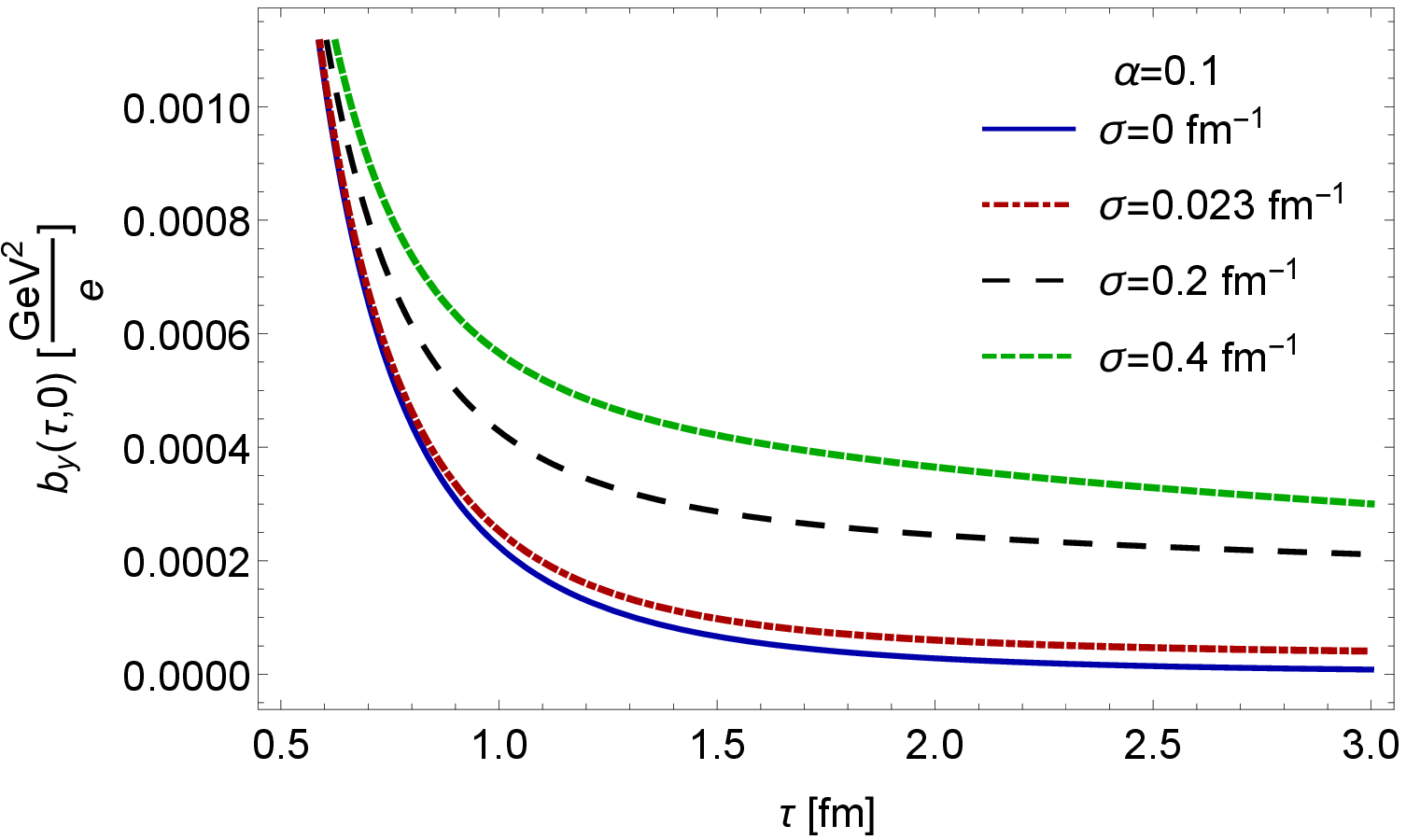}
\caption{}
\end{subfigure}
\\
\begin{subfigure}[b]{0.53\textwidth}
\includegraphics[width=\textwidth]{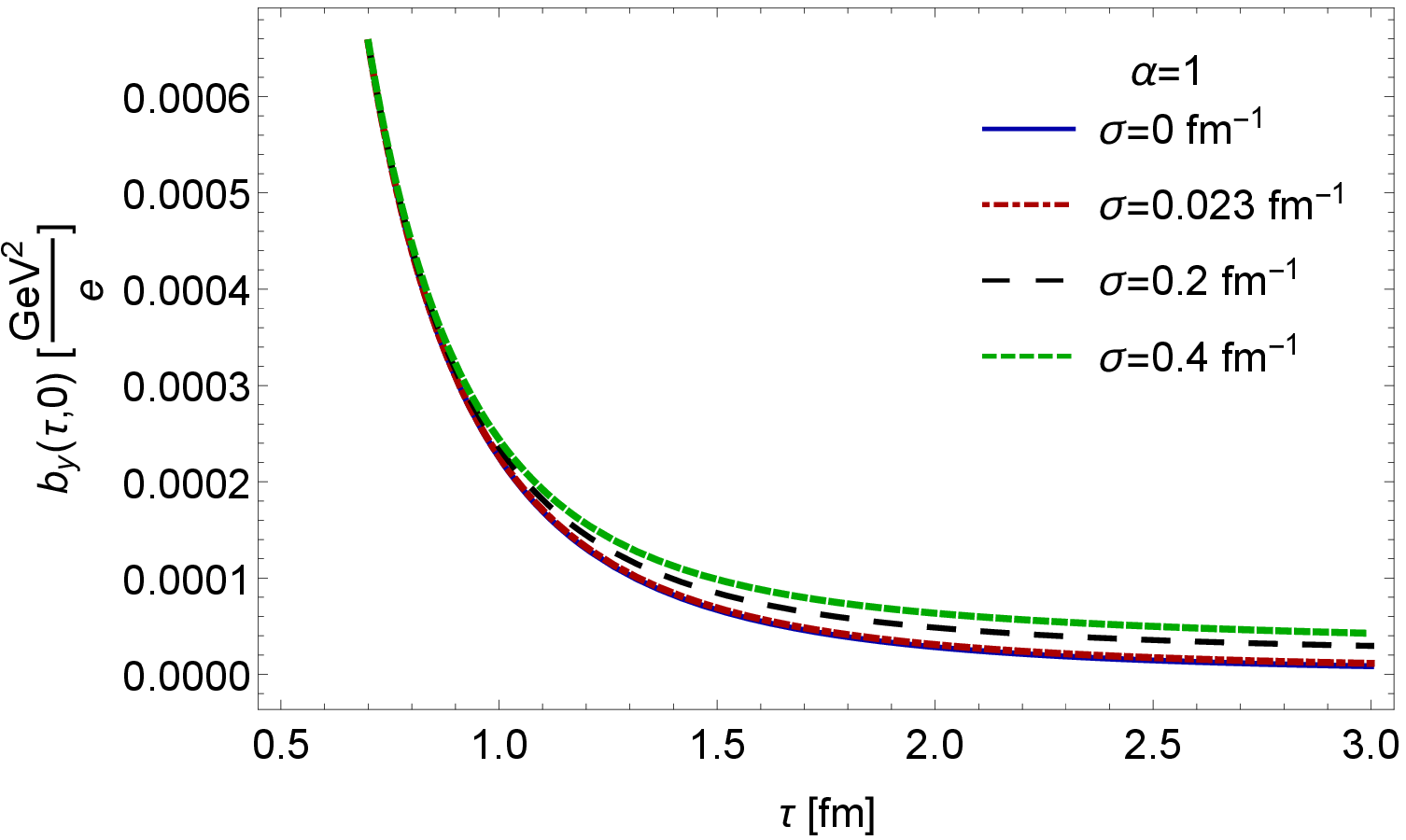}
\caption{}
\end{subfigure}
\begin{subfigure}[b]{0.53\textwidth}
\includegraphics[width=\textwidth]{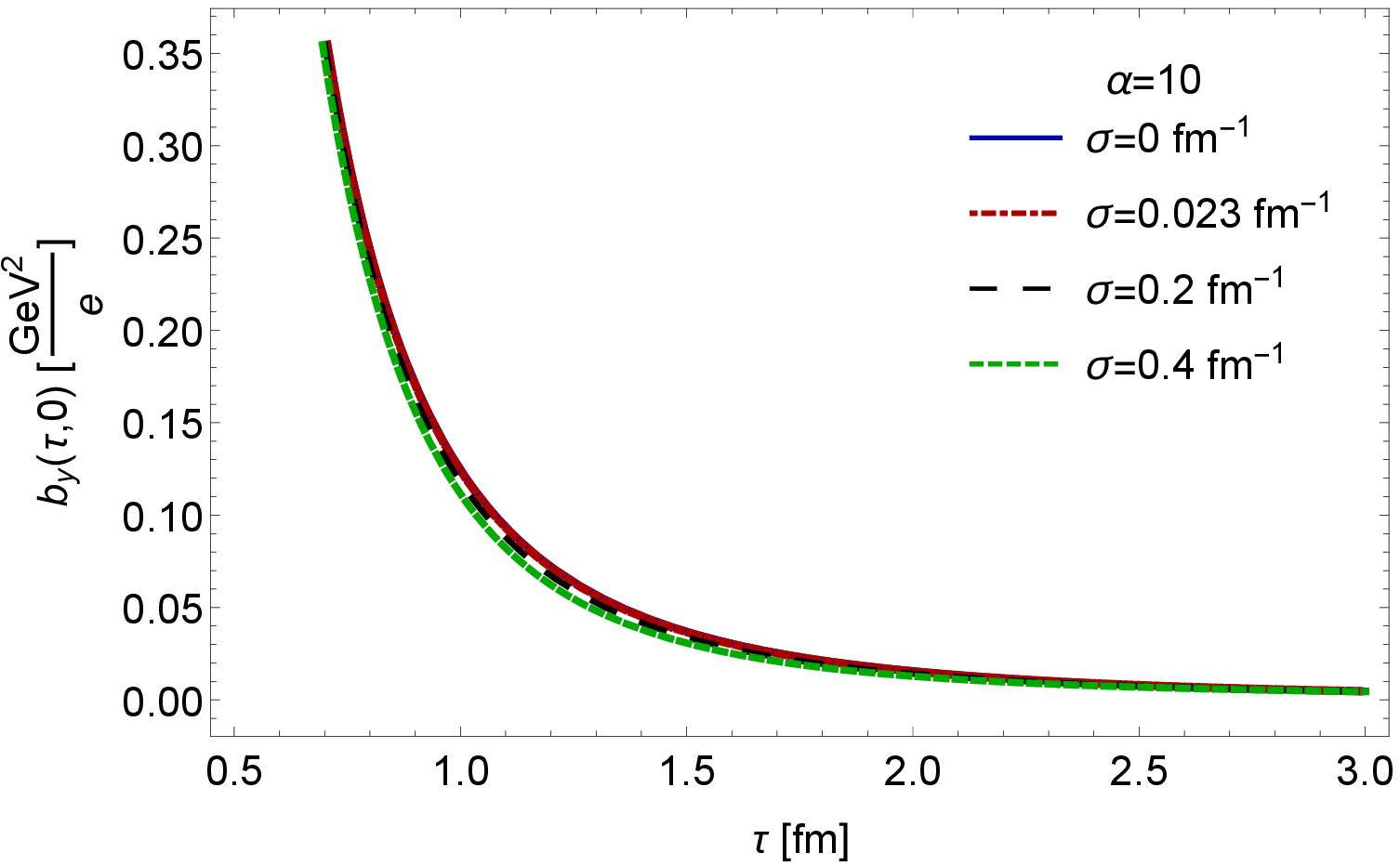}
\caption{}
\end{subfigure}
\caption{\small{The evolution of the magnetic field in term of $\tau$ is plotted for different values of $\alpha$ at $\eta=0$ with $\sigma=0$ (blue solid curve), $\sigma=0.023 \ fm^{-1}$ (red dot-dashed curve), $\sigma=0.2 \ fm^{-1}$ (black  dashed curve), and $\sigma=0.4 \ fm^{-1}$ (green dashed curve). As expected, increasing the electrical conductivity delays  the decay of the magnetic field, which is more notable for $\alpha=0.01$ (Fig.~1(a)), and $\alpha=0.1$ (Fig.~1(b)). }}
\end{figure}

We present a profile of $b_y(\tau,\eta)$ respect to rapidity for different value of  proper times and two  values of  $\sigma$, which is illustrated in Figs.~2(a) and 2(b). Here, each curve refers to different values of proper time from $\tau=0.5\ fm$  to $\tau=3.5 \ fm$. As one can see, the magnetic field increases with rapidity at fix time, and it  tends to decrease with increasing time. Furthermore, the magnetic field tends to remain constant over passing the time at a significant limit of $\eta$. The higher the electrical conductivity, the range of rapidity in which the magnetic field is stable becomes more extensive, and the amount of magnetic field is more valuable too. A comparison between the curves in these figures shows that  the magnetic field increases around the small rapidities with rising the $\sigma$, whereas it decreases in large rapidities.

\begin{figure}
\begin{subfigure}[b]{0.53\textwidth}
\includegraphics[width=\textwidth]{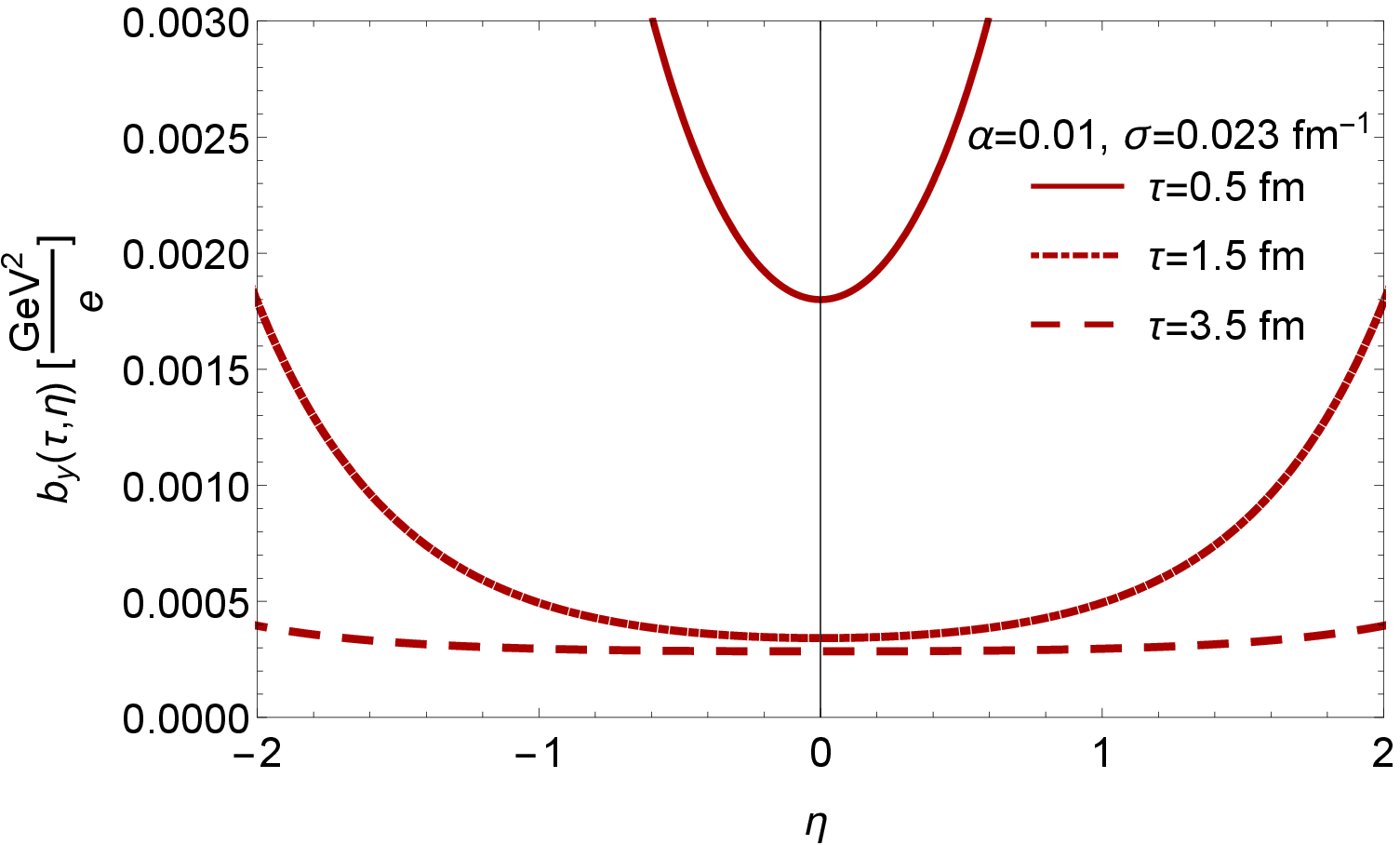}
\caption{}
\end{subfigure}
\hfill
\begin{subfigure}[b]{0.53\textwidth}
\includegraphics[width=\textwidth]{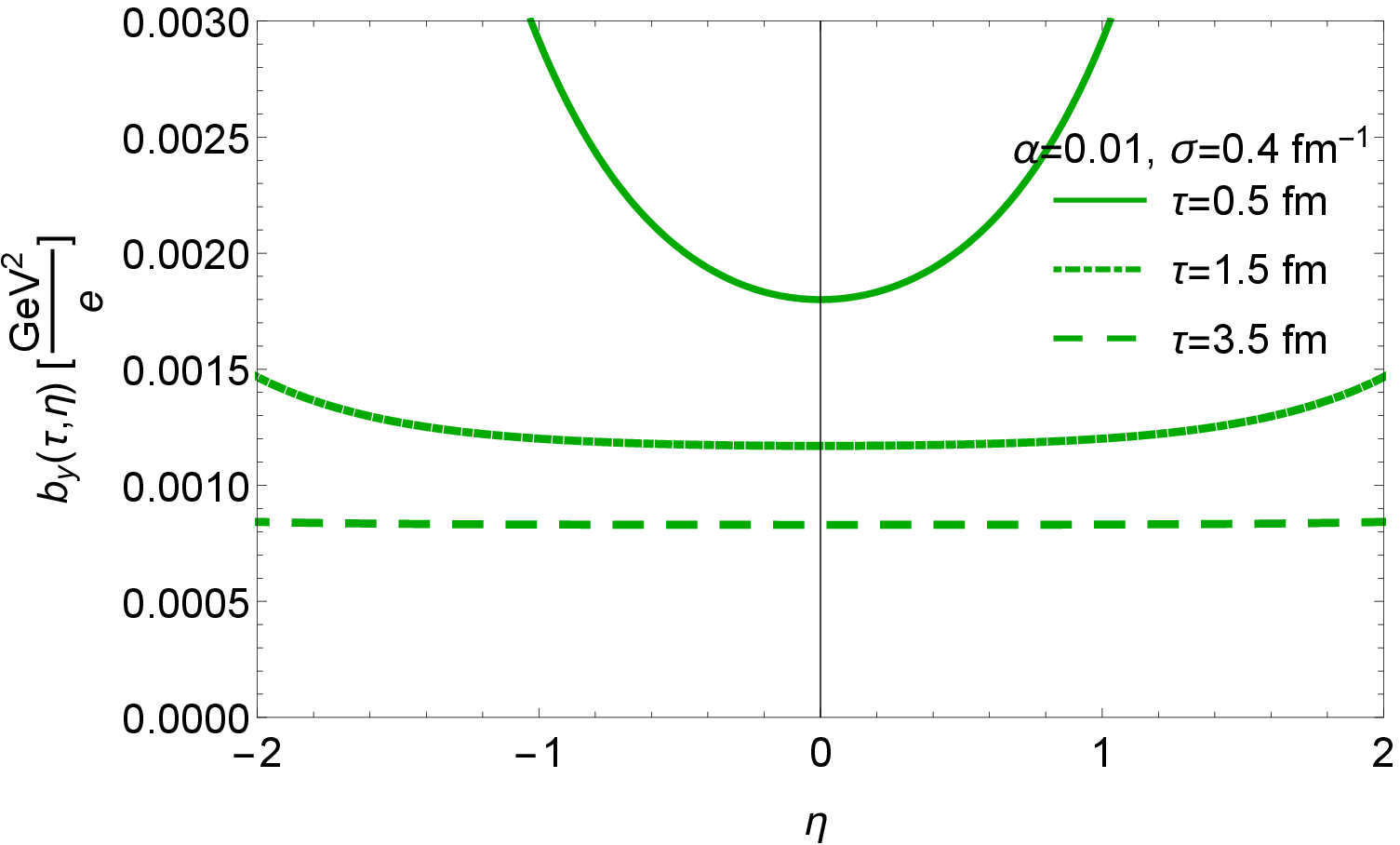}
\caption{}
\end{subfigure}
\caption{\small{The magnetic field in term of  $\eta$ at (a) $\sigma=0.023 \ fm^{-1}$, and (b) $\sigma=0.4 \ fm^{-1}$ is plotted for $\alpha=0.01$. In two panels, the solid curves correspond to $\tau=0.5 \ fm$, the dot-dashed curves to $\tau=1.5 \ fm$, and the dashed curves to $\tau=3.5 \ fm$. The magnetic field remains constant over passing the time at a significant limit, and this range is more inclusive for higher electrical conductivity with the stronger amount of magnetic field (Fig.~2(b)).}}
\end{figure}

In Fig.~3, the time evolution of the electric field  is shown at $\eta=-1$ with different values of the electric conductivity for $\alpha=0.01$ (Fig.~3(a)), and $\alpha=0.1$ (Fig.~3(b)). Similarly to the magnetic field, the electric field decreases  with increasing the  time, and it drops rapidly for the high value of  $\sigma$. It  also shows the electric field has decreased into a faster decay process for smaller value of  $\alpha$ (Fig.~3(a)).  We again consist the value of $\alpha$ depends on initial conditions. Figs.~4(a) and 4(b) are the demonstration of $e_x(\tau,\eta)$ respect to rapidity $\eta$ for different values of the proper time (from $\tau=0.5\ fm$ up to $\tau=3.5 \ fm$) by considering  two different values of $\sigma$. One can see that the electric field is an odd function of $\eta$, and in central rapidity, the electric field is zero. Despite, the electric field is stronger at large rapidities, this amount for $\sigma=0.4 \ fm^{-1}$ (Fig.~4(a)) is smaller than when $\sigma=0.023 \ fm^{-1}$ (Fig.~4(b)); thus, the high conductivity effects on the electric field in fluid, and it will disappear quickly.

\begin{figure}
\begin{subfigure}[b]{0.53\textwidth}
\includegraphics[width=\textwidth]{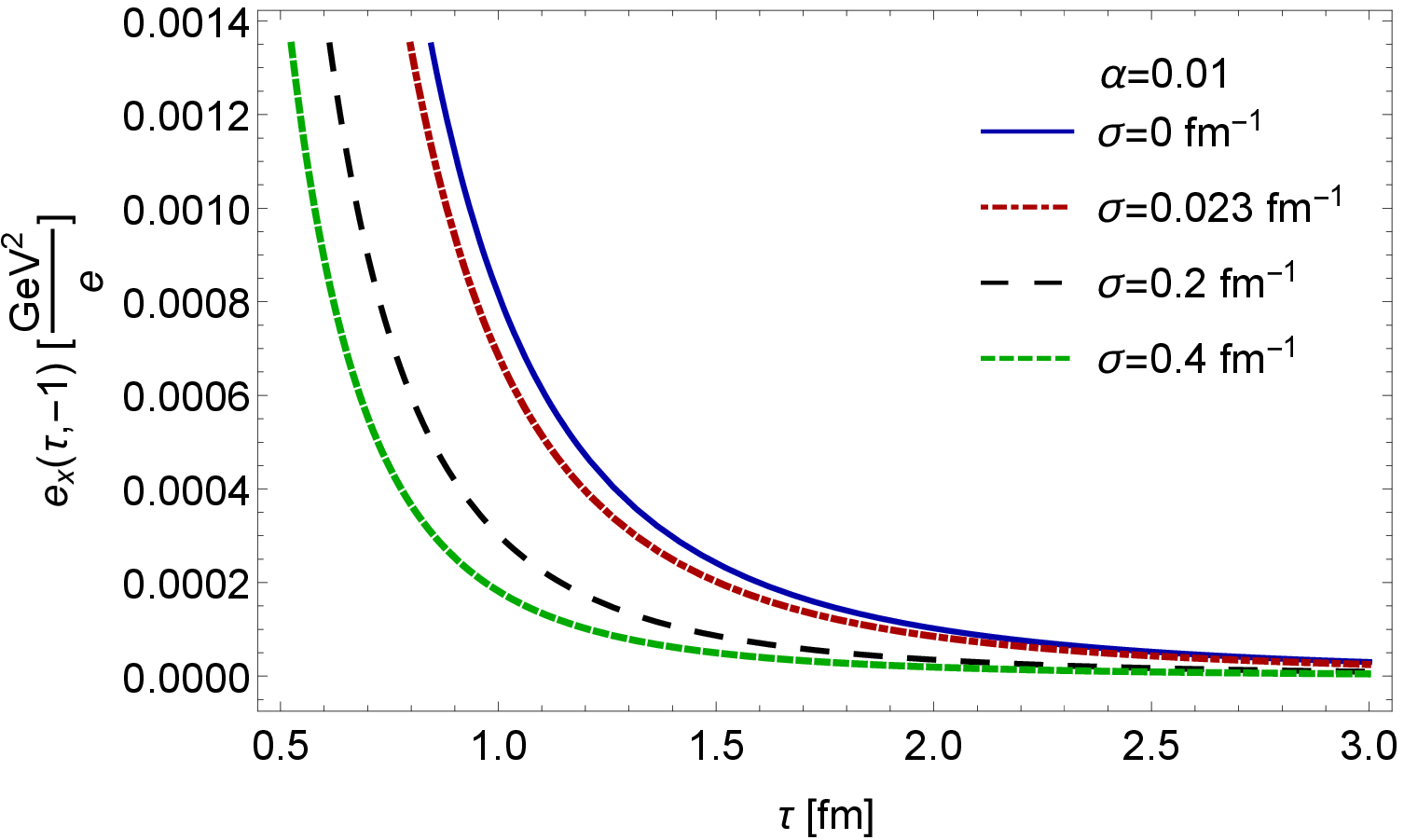}
\caption{}
\end{subfigure}
\hfill
\begin{subfigure}[b]{0.53\textwidth}
\includegraphics[width=\textwidth]{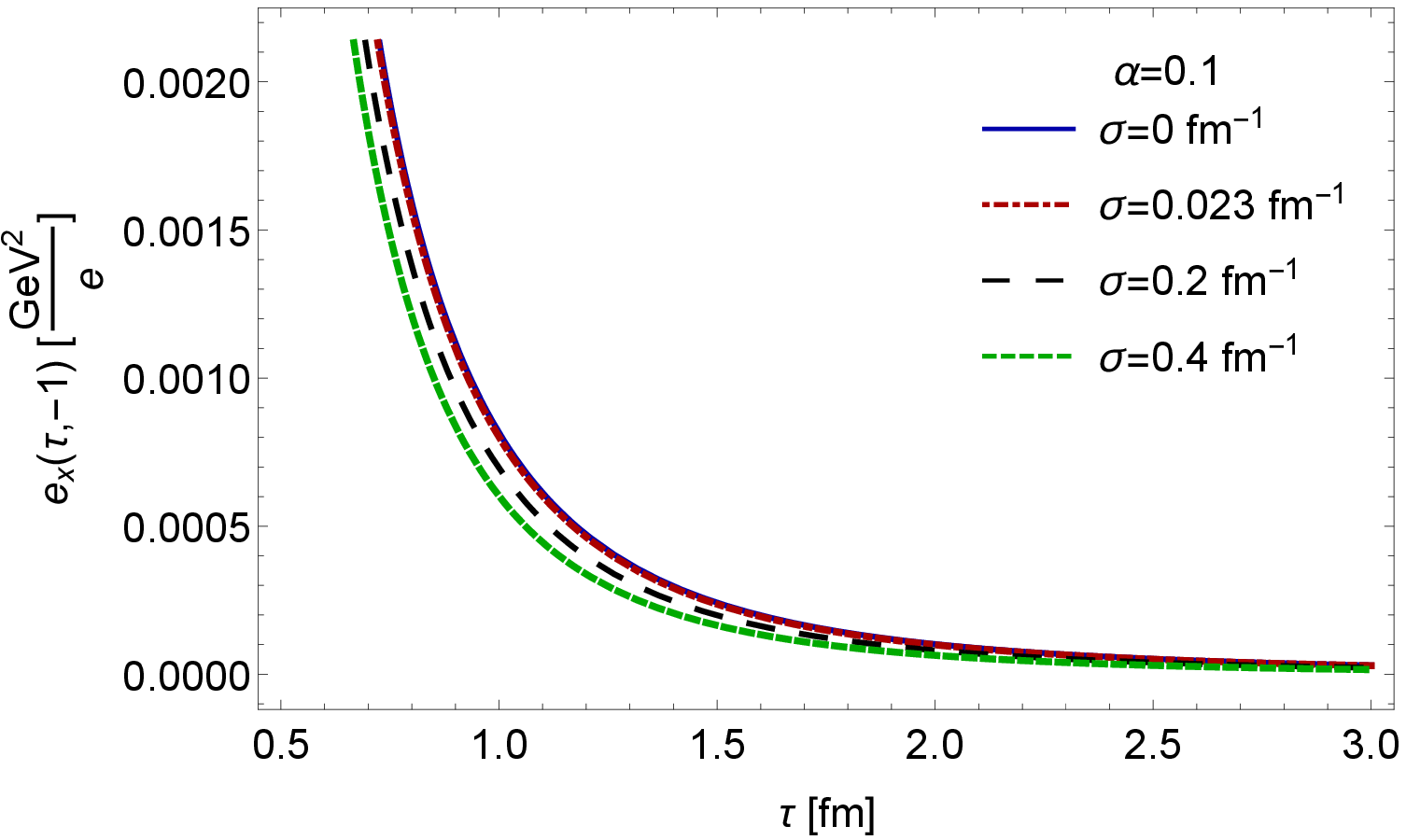}
\caption{}
\end{subfigure}
\caption{\small{The time evolution of $e_x(\tau,\eta)$ is plotted for two different values of $\alpha$ at $\eta=-1$ with $\sigma=0$ (blue solid curve), $\sigma=0.023 \ fm^{-1}$ (red dot-dashed curve), $\sigma=0.2 \ fm^{-1}$ (black dashed curve), and $\sigma=0.4 \ fm^{-1}$ (green dashed curve). As one observes, the electric field decreases with   increasing the electrical conductivity in a quicker process, which is more notable for $\alpha=0.01$ Fig.~3(a).}}
\end{figure}

\begin{figure}
\begin{subfigure}[b]{0.53\textwidth}
\includegraphics[width=\textwidth]{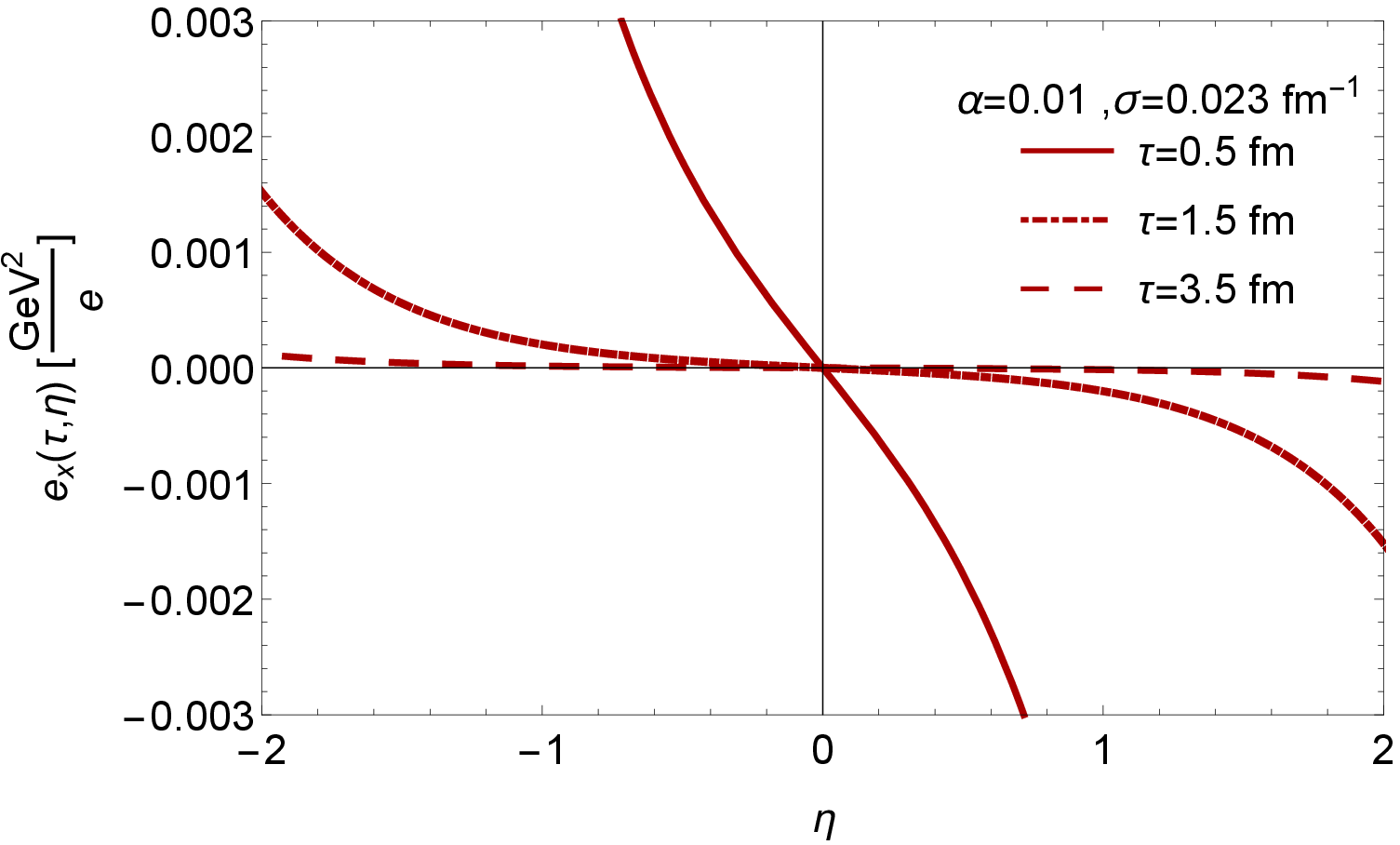}
\caption{}
\end{subfigure}
\hfill
\begin{subfigure}[b]{0.53\textwidth}
\includegraphics[width=\textwidth]{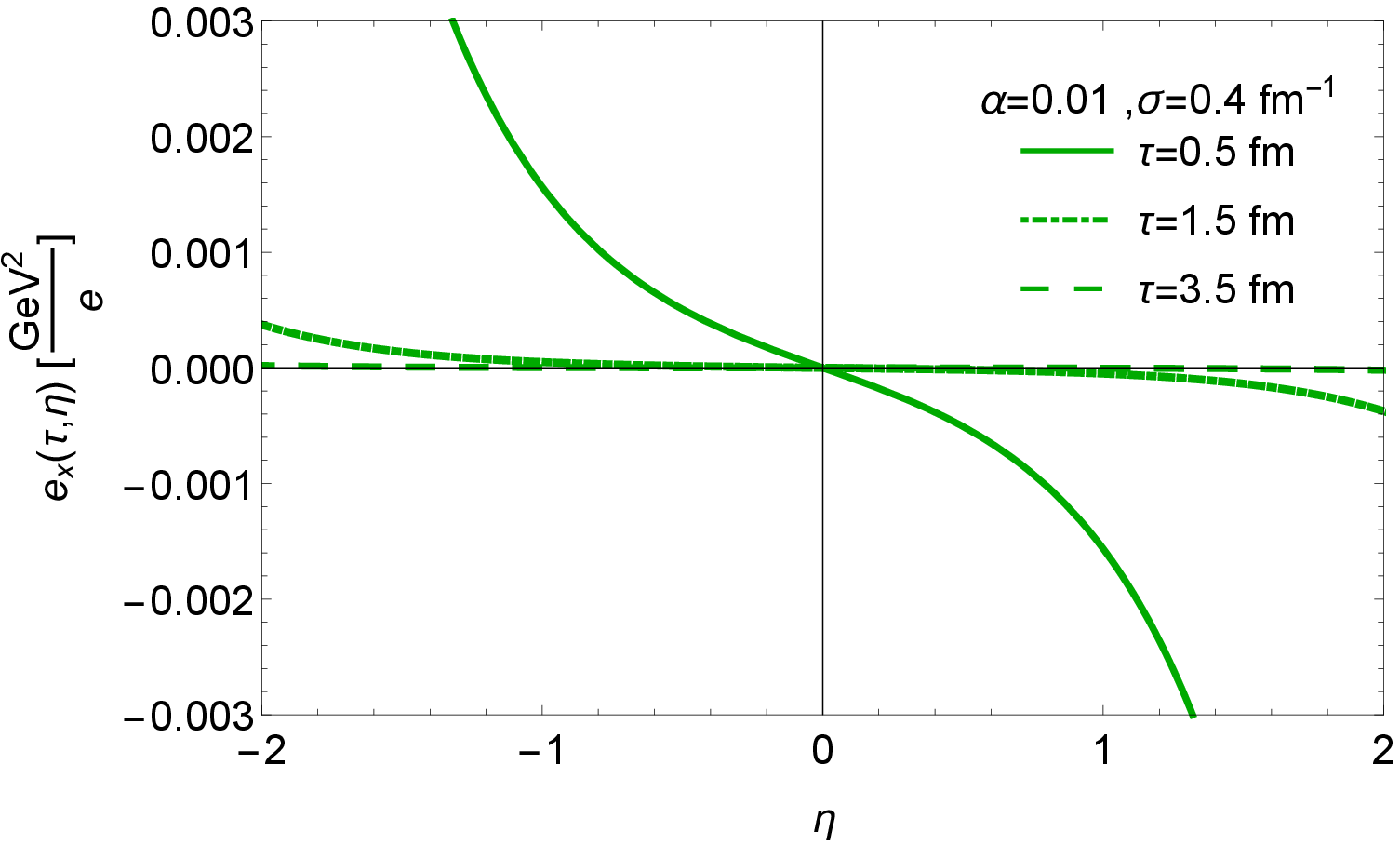}
\caption{}
\end{subfigure}
\caption{\small{The electric field in term of  $\eta$ is outlined at (a) $\sigma=0.023 \ fm^{-1}$, and (b) $\sigma=0.4 \ fm^{-1}$. The value of   $\alpha=0.01$ is chosen. In these plots, the solid curves are related to $\tau=0.5 \ fm$, the dot-dashed curves to $\tau=1.5 \ fm$, and the dashed curves to $\tau=3.5 \ fm$. The electric field is stronger at large rapidities, but the high conductivity causes that the electric field decays quickly in the earlier times.}}
\end{figure}

It is interesting to explore the acceleration parameter, which is obtained from Eq. (\ref{u}). It consists of two critical parts: the effect of generalized Bjorken model (GBM) which illustrates that the acceleration parameter is only dependent on the proper time $\tau$, and the influence of existence of EM fields in the fluid which shows that the acceleration parameter should be considered as a function of $\tau$ and $\eta$.

Fig.~5 is an illustration of evolution of $\lambda(\tau,\eta)$ in term of $\tau$ for different values of rapidity. We choose  the electrical conductivity as $\sigma =0.023 \ fm^{-1}$, and $\alpha=0.01$.  The EM fields with finite electrical conductivity create a considerable acceleration  with a negative sign, especially at the early times, which is more significant at large rapidities and fades the effect of generalized Bjorken model. However, in late time the effect of the EM fields on the acceleration parameter becomes slight, and the acceleration parameter tends to a fixed amount more than 1. Furthermore, the dependence of the acceleration parameter to $\eta$ for fixed $\tau$ is demonstrated in Fig.~6.  It is seen the acceleration parameter decreases when the absolute amount of $\eta$ increases; however, for the late time we have a plateau. According to this plot, increasing  $\eta$ decreases the acceleration parameter, and over passing the time, the reducing process of $\lambda(\tau,\eta)$ is happening in an extended range of rapidity.

\begin{figure}
\begin{center}
\includegraphics[width=9 cm, height=6 cm]{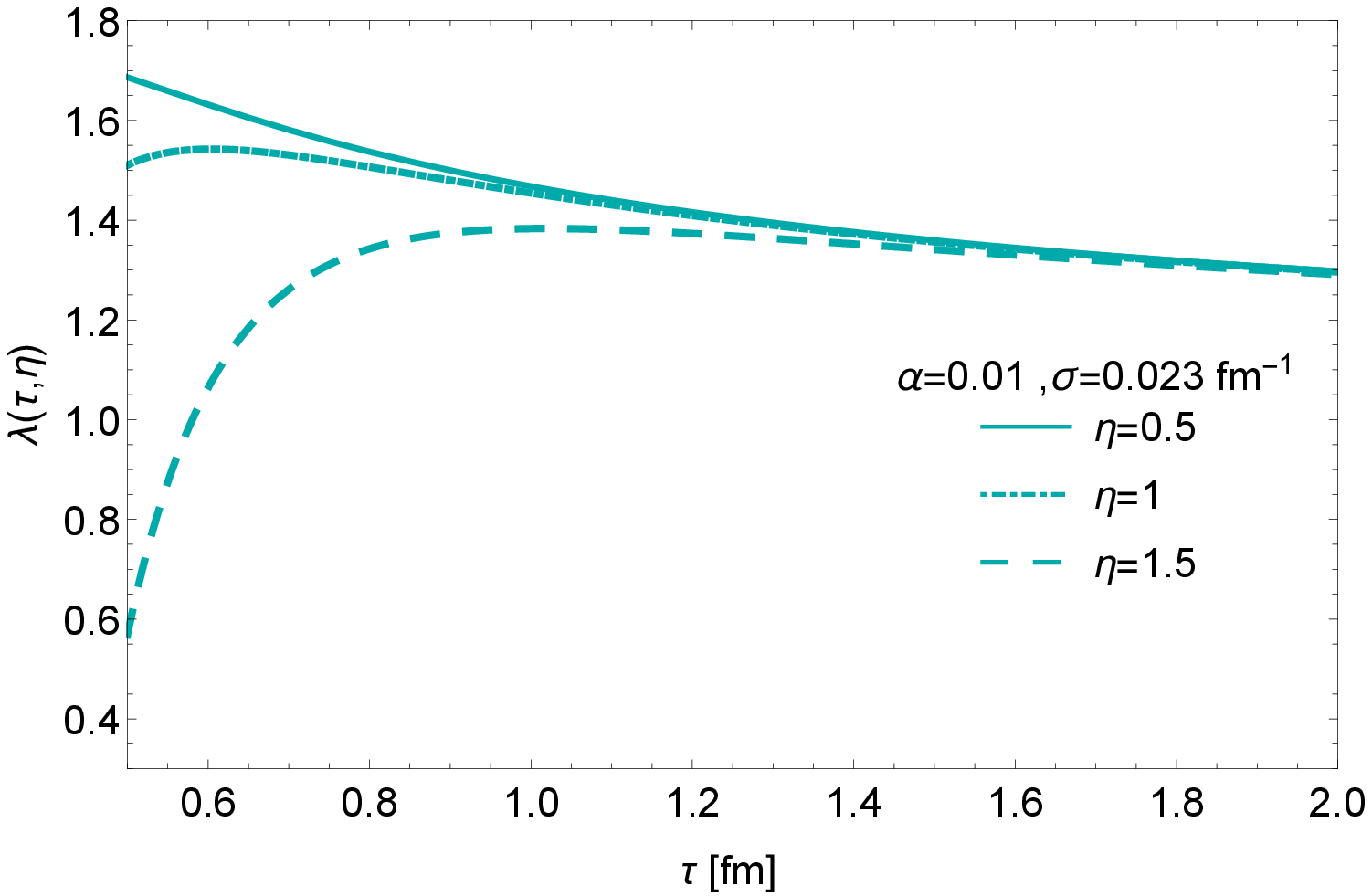}
\caption{\small{Acceleration parameter $\lambda(\tau,\eta)$ in term of proper time $\tau$ is plotted for $\sigma =0.023 \ fm^{-1}$ and $\alpha=0.01$. The solid curve, dot-dashed curve, and the dashed curve corresponds to $\eta=0.5, 1, 1.5$, respectively. As expected, the acceleration parameter tends to a fixed amount more than 1. }}
\end{center}
\end{figure}

\begin{figure}
\begin{center}
\includegraphics[width=9 cm, height=6 cm]{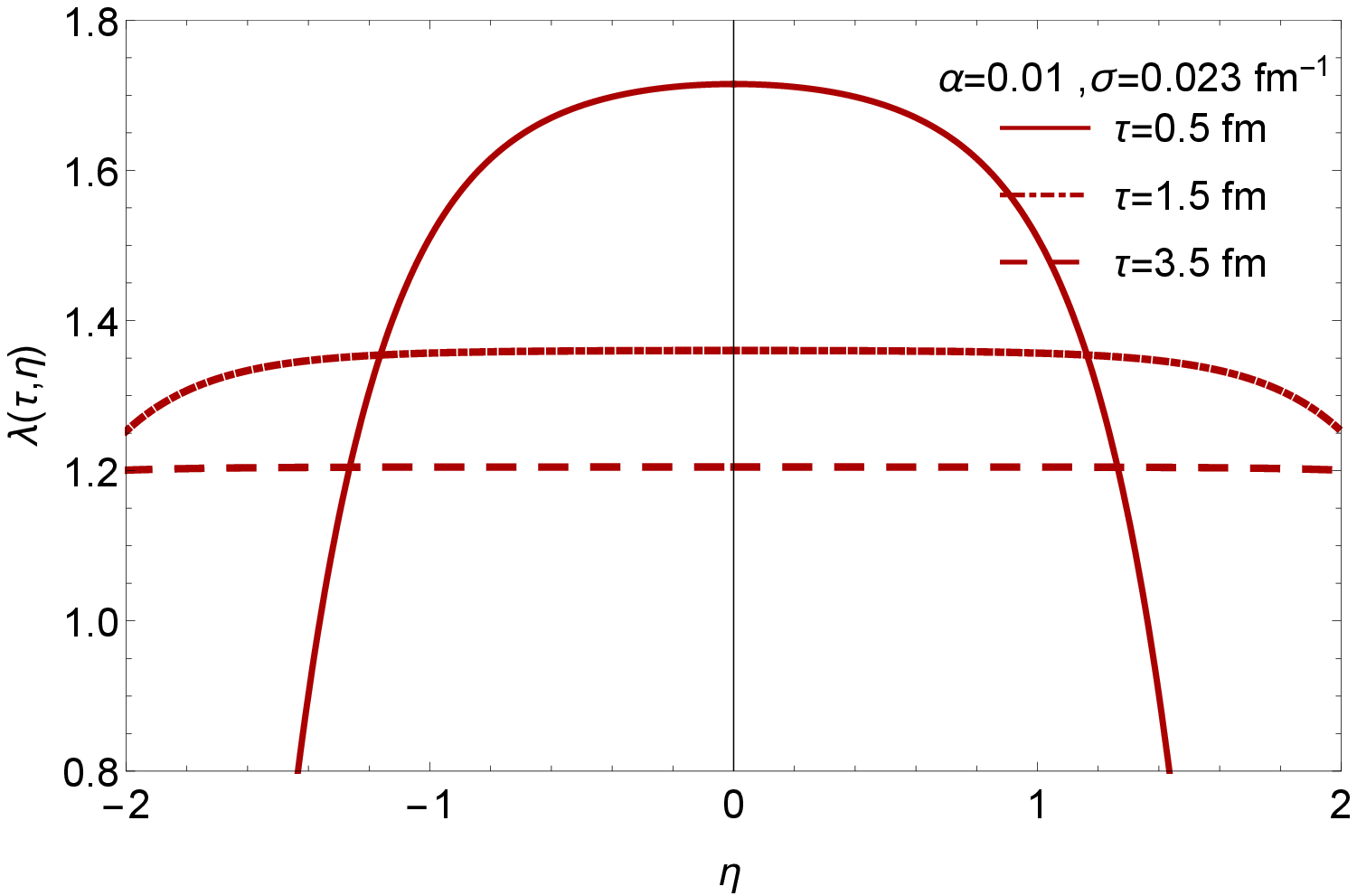}
\caption{\small{Acceleration parameter $\lambda(\tau,\eta)$ in term of  $\eta$ is illustrated for $\sigma =0.023 \ fm^{-1}$, and $\alpha=0.01$ with $\tau=0.5, 1.5 ,3.5 \ fm$ (solid, dot-dashed, and dashed curves, respectively). As it turns out, increasing the $\eta$ decreases the acceleration parameter. Besides at the late time, the reducing process of $\lambda(\tau,\eta)$ is occurring on a broader range of small rapidities.}}
\end{center}
\end{figure}

In order to discuss the dependence of the acceleration parameter on the matter, we investigate the effects of electric conductivity on the $\lambda$, because one of the most important proprieties of the QGP is its electric conductivity.   We know at the mid-rapidity the electric field is zero; therefore, we are not able to investigate the effect of $\sigma$ on the accelerated fluid. Thus, we consider the effect of electrical conductivity $\sigma$ on acceleration parameter $\lambda(\tau,\eta)$ at $\eta=1$. In Fig.~7, the effect of electrical conductivity on acceleration parameter is studied.  Since the electric field  is expected to decay much faster in high conductivity (Fig.~3), the effect of EM fields on the acceleration of fluid is negligible, and $\lambda(\tau,\eta)$ tends to be similar the generalized Bjorken model ($v_z \neq\frac{z}{t}$). The remarkable point in our results is that the acceleration parameter can be found in fluid with high conductivity  (or ideal fluid; $\sigma\to\infty$), too.

\begin{figure}
\begin{center}
\includegraphics[width=9 cm, height=6 cm]{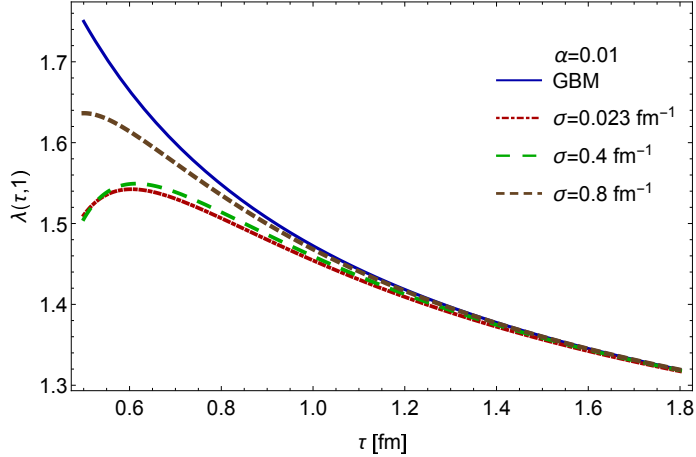}
\caption{\small{The acceleration parameter  $\lambda(\tau,\eta)$  in terms of $\tau$  is indicated for $\eta=1$  and  $\alpha=0.01$. In this plane, the blue solid curve corresponds to generalized Bjorken model (GBM), the red dot-dashed curve to $\sigma =0.023 \ fm^{-1}$, the green  dashed curve to $\sigma =0.4 \ fm^{-1}$, and the brown dashed curve to $\sigma =0.8 \ fm^{-1}$ . As shown, the $\sigma$ influences on the acceleration parameter, and in high conductivity, it leads to generalized Bjorken model.}}
\end{center}
\end{figure}

Finally, we discuss the effects of the EM fields on energy density. One can obtain the correction of energy density by   Eq. (\ref{eeta}). We remind the reader that the total energy density is $\epsilon =\epsilon_0(\tau)+\epsilon_1(\tau,\eta)$ and the latter is the component that is truly affected by the acceleration of the fluid. Here, if we consider the homogeneous solution of Eq.~(\ref{uinh}) as $v^h$ (Eq.~(\ref{vh})), which is the result of generalized Bjorken flow, the correction of energy density will be disappeared. Thus, the energy density is the same as the Bjorken model $ \epsilon_0(\tau)=\epsilon_c (\frac{\tau_0}{\tau})^{1+\kappa}$. Therefore, the correction of energy density is entirely affected by  EM fields and electrical conductivity.

The ratio of energy density $\epsilon(\tau,\eta)/\epsilon_0$ in term of proper time $\tau$ is exhibited in Fig.~8 for $\sigma=0.023 fm^{-1}$, and different values of rapidities. As it is evident, the profile of the energy density is very similar to the acceleration parameter (Fig.~5).  In this plot, we make a comparison between the accelerated fluid and the Bjorken model. We notice that for $\eta=0$ the effect of the EM fields on the energy density could be ignored.  However, the energy density rate decays faster than the Bjorken model at the $\eta \neq 0$, and early time. Also, in Fig.~(9), we demonstrate  $\epsilon(\tau,\eta)/\epsilon_0$ in term of  rapidity for fixed proper times. One can see that at the early times, the plot has  a Gaussian  distribution, while at the late time it becomes a plateau around the small rapidities.
\begin{figure}
\begin{center}
\includegraphics[width=9 cm, height=6 cm]{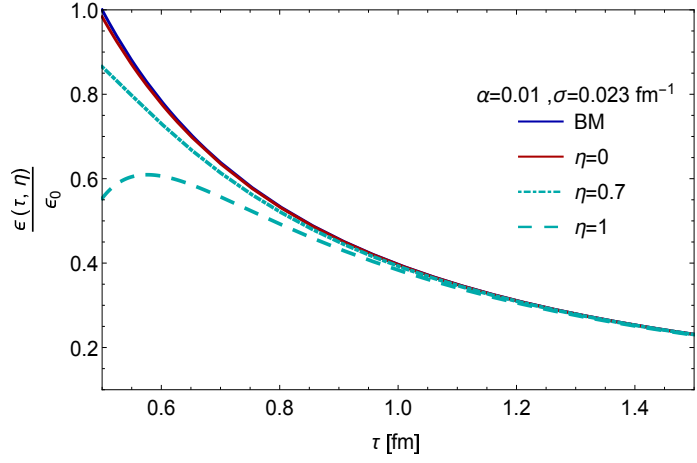}
\caption{\small{The ratio of energy density  $\epsilon(\tau,\eta)/\epsilon_0$ in term of  $\tau$ is outlined for  $\sigma=0.023 \ fm^{-1}$, and $\alpha=0.01$. It is plotted for different rapidities and comparison with Bjorken model. The blue solid curve is related to Bjorken model (BM), the red solid curve to $\eta=0$, the dot-dashed curve to $\eta=0.7$, and the dashed curve to $\eta=1$. }}
\end{center}
\end{figure}

\begin{figure}
\begin{center}
\includegraphics[width=9 cm, height=6 cm]{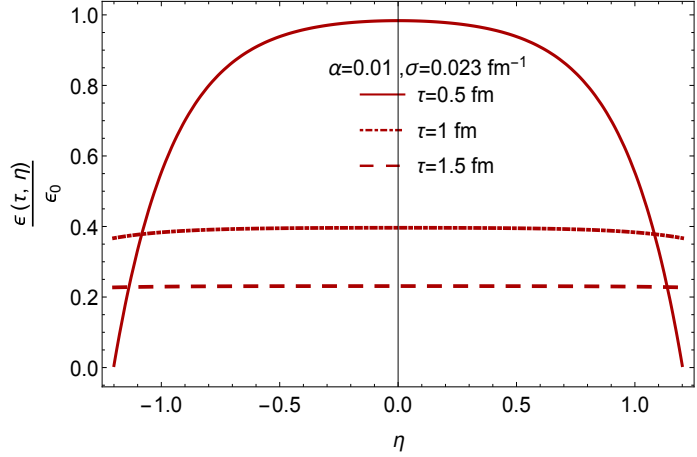}
\caption{\small{The $\epsilon(\tau,\eta)/\epsilon_0$ in term of rapidity  $\eta$ is plotted for  $\sigma=0.023 \ fm^{-1}$, and $\alpha=0.01$ with $\tau=0.5, 1.5, 3.5 \ fm$ (solid, dot-dashed, and dashed curves, respectively). At the late time, the Gaussian form of energy density transforms to a plateau around the small rapidities.}}
\end{center}
\end{figure}

To emphasize the effect of the electrical conductivity on the energy density, we have plotted the time evolution of $\epsilon(\tau,\eta)/\epsilon_0$ at the mid-rapidity by considering several values of $\sigma$. As it is shown in Fig.~10, the reduction of energy density is  modified most significantly at the early times. Due to the rapid decay of the electric field in high conductivity, the influence of $\sigma$ on the perturbation of energy density will be disappeared, and the ratio of energy density tends to the Bjorken model. In addition, $\epsilon(\tau_0,\eta)/\epsilon_0$   in terms of $\eta$ at the fixed proper time $\tau_0=0.5\ fm$ is illustrated in Fig.~11. According to this figure, by increasing the value of the electrical conductivity $\sigma$, the profile of energy density directs to a plateau because in high conductivity the energy density is not sensitive to the rapidity and its behavior leads to the Bjorken model.

\begin{figure}
\begin{center}
\includegraphics[width=9 cm, height=6 cm]{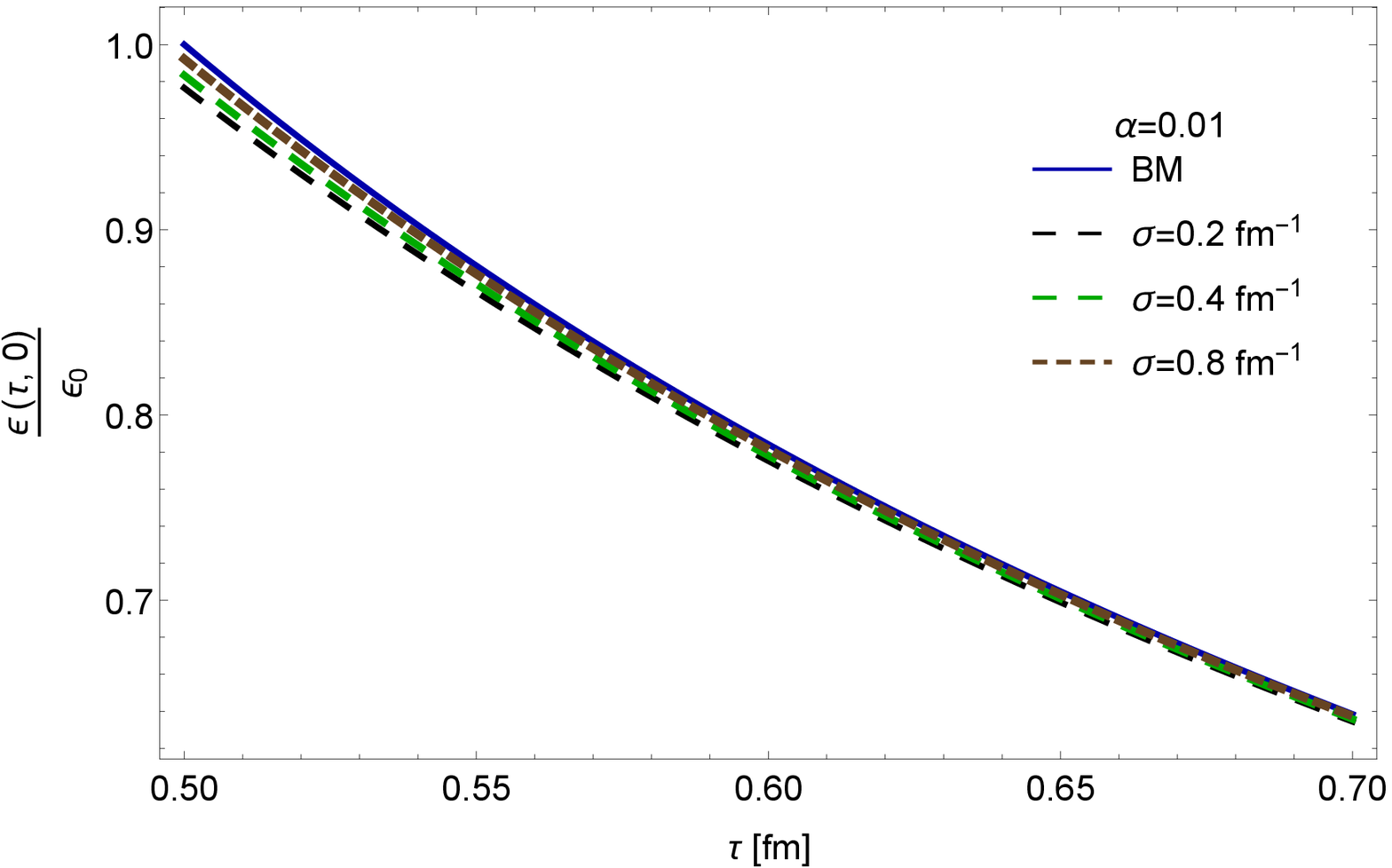}
\caption{\small{The time evolution of the $\epsilon(\tau,\eta)/\epsilon_0$  is plotted for $\alpha=0.01$, and $\eta=0$.  The blue solid curve, the black dashed curve, the green dashed curve, and the brown dashed curve corresponds to the Bjorken model,  $\sigma=0.2\ fm^{-1}$, $\sigma=0.4 \ fm^{-1}$, and $\sigma=0.8 \ fm^{-1}$, respectively. As one observes, increasing the electrical conductivity causes that the decrease of the energy density becomes negligible and leads to the Bjorken model.}}
\end{center}
\end{figure}

\begin{figure}
\begin{center}
\includegraphics[width=9 cm, height=6 cm]{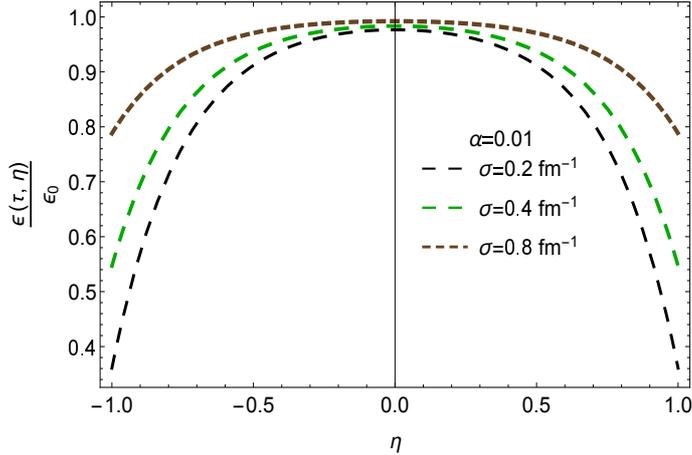}
\caption{\small{The ratio of energy density $\epsilon(\tau,\eta)/\epsilon_0$ in term of  $\eta$ is plotted for $\alpha=0.01$ at early proper time $\tau= 0.5\ fm$.  The black, green, and brown dashed curves correspond to $\sigma=0.2 , 0.4 ,0.8 \ fm^{-1}$,  respectively. The profile of $\epsilon(\tau,\eta)/\epsilon_0$ tends to a plateau in high conductivity. }}
\end{center}
\end{figure}

\subsection{The transverse velocity and transverse momentum spectrum in the presence of electromagnetic fields} 

In the previous sections, we have obtained analytical solutions for the EM fields and energy density. Now we can use these results to estimate the transverse momentum spectrum emerging from the magneto-hydrodynamic solutions. In order to simplify our calculation, we ignore the acceleration effects on the fluid  and assume the  longitudinally boost-invariant expansion at  freeze out temperature. Moreover,
We  assume that the effects of the electromagnetic field are small in the sense that the velocity
of charged particles  resulting 
from the presence of  the electromagnetic field    is much smaller than the velocity of the longitudinal expansion of the plasma.

To examine the effects of electromagnetic fields on the transverse momentum spectrum, we need to obtain the velocity of charged particles such as proton and pion in the spectrum. From $\vec{E}$ and the electrical
conductivity $\sigma$ it would be straightforward to take the electric current density as: $\vec{J}=\sigma\vec{E}$. However, for our purposes, what we need is the transverse velocity which is created by the current density. Besides, we know that when an electric field is established in a conductivity fluid, the charge carriers (assumed positive) acquire a drift speed in the direction of the electric field and this velocity is related to the current density by:  $\vec{J}=nq \vec{v}$, where $n$ is the particle density. In the local fluid rest frame, we look for stationary currents for the $u$ and $d$ quarks and antiquarks. 
 Thus, with setting $q = +2e/3$ for $u$ and $q = +e/3$ for $\bar{d}$, the charge density which is created by the positively charged species obtains as:
\begin{equation}
\vec{J}=\frac{2}{3} n_u\:e \:\vec{v}_u + \frac{1}{3} n_{\bar{d}}\: e \: \vec{v}_{\bar{d}}=en(\frac{2}{3} \:\vec{v}_u + \frac{1}{3}\: \vec{v}_{\bar{d}})=en\vec{v}
\end{equation}
For simplicity, we assume that the particle density for  quarks and antiquarks are the same, and these particles have a similar share in the transverse velocity of the fluid. With these assumptions, the transverse velocity becomes:
\begin{equation}
v_x = \frac{\sigma e_x}{ne}
\end{equation}
where $e_x$ can be taken from Eq.~(\ref{eb}). 
\\
\\
In order to investigate the effects of the conductivity of fluid on the hadron spectra, we  calculate the hadron spectra for the pions and the protons.  we follow the standard prescription which
was developed by Cooper and Frye. 
 
The final transverse spectrum is calculated at the freeze-out surface via the Cooper-Frye(CF) formula is given by \cite{b24,b36}:
\begin{eqnarray}\label{SP}
S= \frac{g_i}{2\pi} \int ^{x_f} _0 m_T x_\perp \tau_f(x_\perp) K_1\left( \frac{m_T u_{\tau}}{T_f}\right) I_0\left(\frac{m_T u_{x_\perp}}{T_f} \right) dx_\perp 
\end{eqnarray}
where $u_{\tau}=1$ , $u_{x_\perp}=\gamma v_x=v_x$, and $\tau_f(x_\perp)$ is the solution of the $T(\tau_f, x_\perp)=T_f$, where $T_f$ is the temperature at the freeze-out surface. It is the isothermal surface in space-time at which the temperature of the inviscid  fluid is related to the energy density as $T\propto \epsilon ^{1/4}$ and for the energy density, the relation $\epsilon =\epsilon_0(\tau)+\epsilon_1(\tau,\eta)$ is applied where $\epsilon_1(\tau,\eta)$ is the correction of energy density and is obtained from Eq.~(\ref{eeta}). Also, $p_T$ is the detected transverse momentum, $m_T=\sqrt{m^2+p_T^2}$ the corresponding transverse mass and $g_i=2$ is degeneracy factor for the proton or pion.

The transverse spectrum Eq. (\ref{SP}) is illustrated in Fig.~12 and Fig.~13 for three different values of electrical conductivity and compared with experimental results obtained at PHENIX \cite{b37} in non-central Au-Au collisions. Here, we consider $T_f=130 MeV$. Our spectrum obtained from the above calculations is not in a good agreement with experimental data,  but  appears to underestimate the experimental data, and their behavior with $p_T$ has the correct trend of a monotonically decrease. Although the conductivity has a significant effect on the electric field, magnetic field and energy density, the final spectrum is not sensitive to $\sigma$ parameter.
\begin{figure} 
\centerline{\begin{tabular}{cc}
\includegraphics[width=12 cm, height=7 cm]{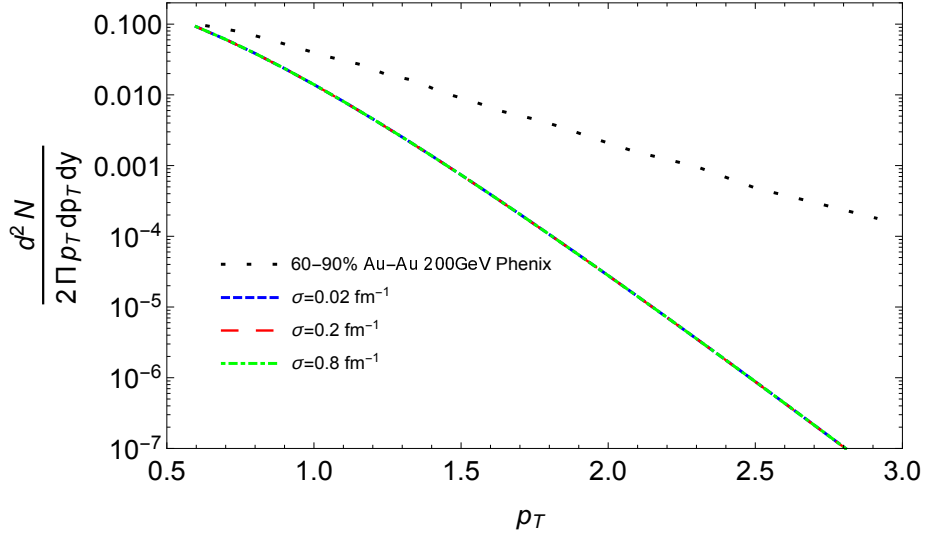}
\end{tabular}}
\caption{The Proton transverse spectrum from non-central Au-Au collisions. The red, black, and brown curves correspond to $\sigma=0.023, 0.2, 0.8 \ fm^{-1}$, respectively, and the blue curve corresponds to PHENIX data.  }
\end{figure}

\begin{figure} 
\centerline{\begin{tabular}{cc}
\includegraphics[width=12 cm, height=7 cm]{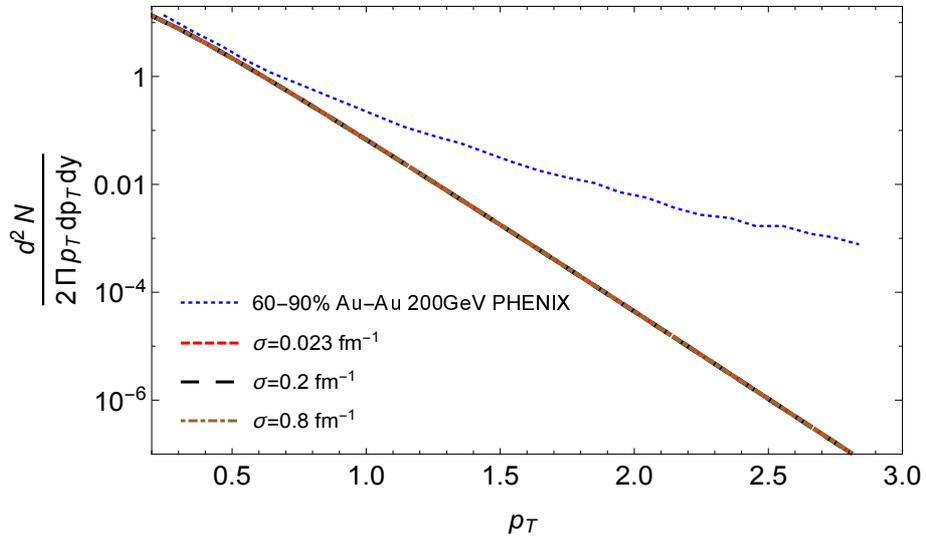}
\end{tabular}}
\caption{The Pion transverse spectrum from non-central Au-Au collisions. The red, black, and brown curves correspond to $\sigma=0.023, 0.2, 0.8 \ fm^{-1}$, respectively, and the blue curve corresponds to PHENIX data. }
\end{figure}

\section{Conclusion and outlook}

The magnitudes and the evolution of electromagnetic fields play a crucial role in estimations of possible observable effects of the de-confinement and chiral phase transitions in heavy-ion collisions. There have been many works that have investigated the electromagnetic field  strength and their evolution. It has been known
in the initial stage, the magnitude of the magnetic field falls rapidly with time ($|B_y|\sim \frac{1}{\tau^3}$); however, the presence of the hot quark-gluon plasma (QGP) may increase the lifetime of the strong magnetic field.

In the present work, we have studied  1+1 longitudinal acceleration expansion motion of a fluid with the electrical conductivity $\sigma$ in the magneto-hydrodynamic framework in the presence of EM fields. Making use of Milne coordinates, in our setup, the magnetic field pointed in $y$ and the electric field pointed in $x$ directions. We assume the Bjorken form of the fluid velocity in the longitudinal direction and  the effect of the electromagnetic field on the longitudinal expansion of the fluid  as a perturbation. This perturbation results in a correction in the energy density of the fluid. Since we are interested in the longitudinal expansion, all quantities depended on proper time $\tau$ and space-time rapidity $\eta$. By applying the initial conditions in Maxwell's equations, we  computed  analytical solutions for the electric and magnetic fields in the conducting fluid with the electrical conductivity $\sigma$, and the energy and Euler equations provided us an accurate description of the rapidity of fluid and the perturbation of the energy density.

We have shown the time evolution of the magnetic and electric fields respect to the $\tau$ and $\eta$ variables and the effect of $\sigma$ parameter in the evolution. The analytical results are in a satisfactory agreement with the previous ones, which in the ideal RMHD is considered, and the electrical conductivity is infinite. Besides, we found a picture of the evolution of the rapidity of fluid and energy density. Our analytic solutions are worth in order to acquire deeper understandings for  more  realistic numerical results.

\end{document}